\documentclass[prd,aps,amsmath,amssymb,nofootinbib,twocolumn,preprintnumbers]{revtex4}

\usepackage{epsfig}
\usepackage{amssymb}
\usepackage{bm}
\usepackage{hyperref}
\usepackage{color}
\usepackage{amsmath}

\newcommand{\be}{\begin{equation}}
\newcommand{\ee}{\end{equation}}
\newcommand{\bea}{\begin{eqnarray}}
\newcommand{\eea}{\end{eqnarray}}

\newcommand{\nn}{\nonumber}

\newcommand{\degree}{^\circ}

\def\lsim{\:\raisebox{-0.75ex}{$\stackrel{\textstyle<}{\sim}$}\:}

\begin{document}

MI-TH-1632, WSU-HEP-1608, LA-UR-16-29252

\title{\bf\Large Indirect Signals from Solar Dark Matter Annihilation to Long-lived Right-handed Neutrinos}

\author{Rouzbeh Allahverdi$^1$}
\author{Yu Gao$^{2,3}$}
\author{Bradley Knockel$^1$}
\author{Shashank Shalgar$^4$}
\affiliation{\bigskip 
$^{1}$~Department of Physics and Astronomy, University of New Mexico, Albuquerque, NM 87131, USA 
$^{2}$~Mitchell Institute for Fundamental Physics and Astronomy, \\
Department of Physics and Astronomy, Texas A\&M University, \\ College Station, TX 77843-4242, USA \\
$^{3}$~Department of Physics and Astronomy, Wayne State University, Detroit, MI 48201, USA \\
$^{4}$ Theoretical Division, Los Alamos National Laboratory, Los Alamos, New Mexico 87545, USA
}

\begin{abstract}
We study indirect detection signals from solar annihilation of dark matter (DM) particles into light right-handed (RH) neutrinos with a mass in a $1-5$ GeV range. These RH neutrinos can have a sufficiently long lifetime to allow them to decay outside the Sun and their delayed decays can result in a signal in gamma rays from the otherwise `dark' solar direction, and also a neutrino signal that is not suppressed by the interactions with solar medium. We find that the latest Fermi-LAT and IceCube results place limits on the gamma ray and neutrino signals, respectively. Combined photon and neutrino bounds can constrain the spin-independent DM-nucleon elastic scattering cross section better than direct detection experiments for DM masses from 200 GeV up to several TeV. The bounds on spin-dependent scattering are also much tighter than the strongest limits from direct detection experiments.   
\end{abstract}

\maketitle

\section{Introduction}
\label{sect:intro}

Signals from dark matter (DM) annihilations inside the Sun~\cite{bib:sun} have been extensively studied in the context of DM indirect detection searches. DM particles, upon scattering off solar medium, can become gravitationally trapped and start annihilating into standard model (SM) particles after their numbers build up at the center of the Sun. Neutrinos thus produced can escape from the Sun and those with energies above weak scale can be detected by active Cherenkov detectors like IceCube~\cite{ICBounds} and Antares~\cite{Adrian-Martinez:2016gti}. Low-energy neutrinos from stopped pions
may also be used as a probe of DM annihilation inside the Sun~\cite{Low}. Provided that equilibrium between the capture and annihilation of DM particles inside the Sun is established, the flux of neutrinos is determined by the capture rate~\cite{JKG}. It can therefore be used to constrain the DM-nucleon elastic scattering cross section. The bounds thus set are much tighter than those form direct detection experiments for spin-dependent (SD) interactions~\cite{ICBounds}, while for spin-independent (SI) interactions 
direct detection experiments, like LUX~\cite{LUX} and PandaX~\cite{PandaX}, often set much stronger limits.

DM annihilation inside the Sun may also result in a photon signal in case that it produces relatively long-lived intermediates states that can escape from the Sun before decaying. Such a scenario can arise in various new physics models, with the dark photon~\cite{Schuster:2009au}, secluded~\cite{Batell:2009zp}, inelastic~\cite{Menon:2009qj}, boosted~\cite{Berger:2014sqa}, portal~\cite{Meade:2009mu} DM models as examples. Since the solar direction is dark in high energy cosmic gamma rays, any photonic signal in that direction offers a stringent constraint.

DM particles may annihilate to the right-handed (RH) neutrinos, which in turn decay to the SM particles, at a significant rate in simple extensions of the SM. A minimal and well-motivated example is the supersymmetric extension of the SM that includes a gauged $U(1)_{B-L}$ symmetry~\cite{MM} (where $B$ and $L$ are baryon number and lepton number respectively). Anomaly cancellation then implies the existence of three RH neutrinos and allows us to write the Dirac and Majorana mass terms for the neutrinos to explain the mass and mixing of the light neutrinos. This model provides two new DM candidates, the lightest neutralino in the $B-L$ sector and the lightest RH sneutrino, both of which may dominantly annihilate into the RH neutrinos. The decay of these RH neutrinos can then lead to interesting indirect detection signals~\cite{ABDR,ACD,ACDG}.

RH neutrinos with a mass much below the weak scale undergo three-body decay via off-shell $W$ and $Z$ bosons through their small mixing with the left-handed (LH) neutrinos.
For masses in the 1-few GeV range, RH neutrinos produced from solar DM annihilation can readily obtain a long ($\sim$1-10 seconds) lifetime. This implies that a significant fraction of the RH neutrinos can decay outside the Sun,  resulting in distinct neutrino and photon signals compared with the usual scenario where DM annihilation produces SM particles inside the Sun. These signals can be used to limit the DM-nucleon elastic scattering cross sections. As we will see, the Fermi-LAT and IceCube data together can constrain the SI cross sections better than direct detection experiments for DM masses in the 200 GeV to 5 TeV range.

In this paper, we adopt a model-independent approach to study the neutrino and photon signals from solar DM annihilation into long-lived RH neutrinos.  In Section~\ref{sect:model}, we briefly discuss the case for light RH neutrinos. As we discuss in Section~\ref{sect:spectra}, depending on its mass, long-lived RH neutrinos can yield characteristic spectra in the neutrino and gamma ray signals. In Section~\ref{sect:indirect}, we perform an analysis of the signals at IceCube and Fermi-LAT and obtain constraints in the parameter space consisting of the RH neutrino mass and DM mass. Finally, we conclude the paper and discuss future prospects in Section~\ref{sect:conclusion}.

\section{Long-lived right-handed neutrinos}
\label{sect:model}

In the simplest case of Type-I seesaw~\cite{bib:typeI}, the RH neutrino $N$ with an arbitrary Majorana mass $M_N$ mix with the SM neutrinos through a Yukawa term as follows
\be 
\Delta {\cal L} \supset y_{\rm D} (L^{\dagger} \cdot i\tau_2 H) N + {\bf h.c.} ,
\ee
where we ignore flavor indices for simplicity. After the Higgs field acquires a vacuum expectation value (vev), the Yukawa term induces a mixing  $\theta \sim \frac{y_{\rm D}}{M_N} v$ between $N$ and the LH neutrino $\nu$. We choose the nominal value for the mixing such that it gives rise to the light neutrino mass $m_\nu$:
\be \label{nominal}
\theta \approx \left({m_\nu \over M_N}\right)^{1/2} .
\ee
The mass of each light mass eigenstate $m_\nu$ receives contributions from mixing with the three RH neutrinos. Hence, for a given RH neutrino, $\theta$ may be larger or smaller than the nominal value in above. In the former case, mixings from different RH neutrinos must cancel out to give the right value of $m_\nu$. In the latter case, the other RH neutrinos should make the main contribution to $m_\nu$.  

Although the magnitude of $M_N$ is often assumed to be much larger than the electroweak scale, some or all of $N$'s can have a mass around or below the electroweak scale. This, for example, can happen in the split seesaw scenario~\cite{split}. Values of $M_N$ around the electroweak scale are phenomenologically very interesting as they provide an opportunity for experimental discovery of the RH neutrinos thereby potentially unveiling the mechanism
of neutrino mass generation. Masses up to 500 GeV are accessible at the LHC~\cite{LHC}, and the prospect would be even better at a high-energy lepton collider~\cite{LC}.

RH neutrinos with a mass $M_N < 5$ GeV bring in new possibilities to test them experimentally. They can be searched for in meson decays at $B$ and $K$ factories~\cite{Meson}, fixed target experiments~\cite{FT}, and the SHiP experiment~\cite{SHIP} proposed at CERN. The implications of such light RH neutrinos for the neutrinoless double-beta decay have also been studied~\cite{0nu2beta}. Here we focus on RH neutrinos with a mass in the $\sim 1-5$ GeV range.

Given the eV~\cite{bib:pdg2016} or sub-eV~\cite{bib:cosmo} scale of current neutrino mass limits, heavy neutrinos of a GeV scale mass would imply a nominal mixing $\theta \sim 10^{-6}-10^{-5}$, see Eq.~(\ref{nominal}). Since this mixing is generally small, from now on we use $N$ to denote the heavy mass eigenstate after the mixing.

$N$ can decay into to SM neutrinos and the Higgs via its dominant singlet component, as well as SM gauge bosons and leptons via its small mixing with LH neutrinos. If $M_N > m_H$, the Higgs decay channel will dominate\footnote{Indirect detection signals of DM annihilation to RH neutrinos for this case have been studied in the context of the supersymmetric $U(1)_{B-L}$ extension of the SM~\cite{ABDR,ACD,ACDG}.}. For $m_{W,Z} < M_N < m_H$, the gauge boson channels are dominant. In the case that $M_N < m_{W,Z}$, as we consider here, three-body decays via off-shell $W$ and $Z$ bosons will be the main channels\footnote{Three-body decays via off-shell Higgs are subdominant due to small Yukawa couplings of the Higgs to fermions.}.   
The $N$ decay is then dominated by the weak interaction from its SM lepton component, resulting in the following boosted decay width: 
\be \label{width}
\Gamma_N \propto \theta^2 G^2_{\rm F} M^5_N {M_N \over M_{\rm DM}}.
\ee
Here $G_{\rm F} M^5_N$ is the rest frame decay width (up to a phase space factor), $M_N/M_{\rm DM}$ is due to the Lorentz boost, and $\theta$ is the $N-\nu$ mixing. The detailed expression for the partial widths of leptonic and semi-leptonic decay modes of $N$ are given in~\cite{Decay}. After using the nominal value for mixing in Eq.~(\ref{nominal}), the decay lifetime is found to be:
\be \label{lifetime}
\tau_N \propto {M_{\rm DM} m_\nu \over M^5_N}.
\ee

\begin{figure}
\includegraphics[width=0.45\textwidth]{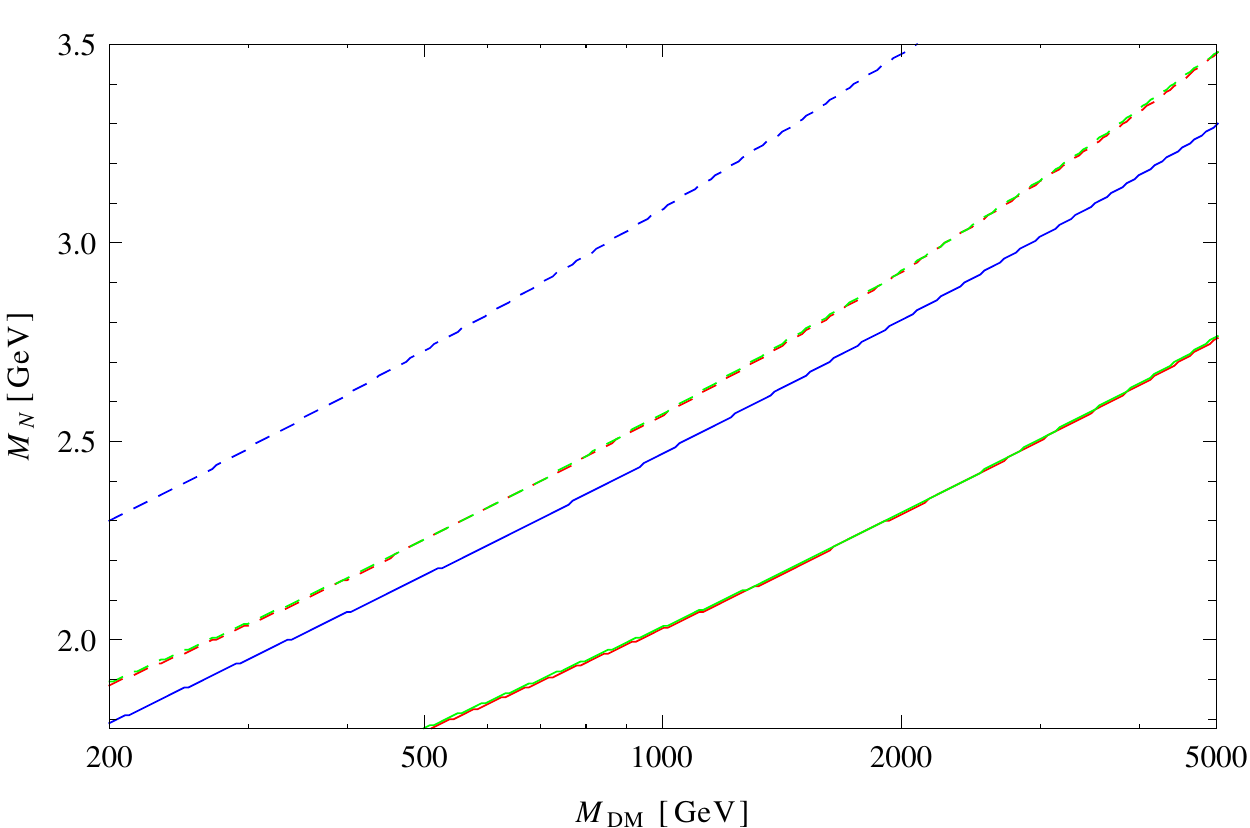}
\caption{The characteristic decay length is inside the photosphere (above the solid lines) and is within 200,000 km (above the dashed lines). The colors red, green, and blue correspond to the cases when $N$ mixes dominantly with $\nu_e$, $\nu_\mu$ and $\nu_\tau$ respectively. The red and green curves almost coincide.}
\label{fig:lifetimes}
\end{figure}

We discuss the details of calculating $\tau_N$ for light RH neutrinos in the next Section. In Fig.~\ref{fig:lifetimes}, we show the $N$ lifetime contours in the $M_N-M_{\rm DM}$ plane that correspond to two characteristics decay lengths: the Sun's photosphere $R_{\odot} \approx 700,000$ km, and an `escape' $R \approx 200,000$ km for neutrinos with less than TeV energy. RH neutrino decays outside the photosphere $R_{\odot}$ give rise to a photon signal. On the other hand, decays happening outside the $200,000$ km radius $R$ produce neutrinos that propagate largely unaffected by interactions with the solar medium, while the associated photons are absorbed. We have checked that neutrinos with energy $E_\nu \lsim 1$ TeV that are produced at distances larger than this experience less than 10$\%$ attenuation before completely leaving the Sun (more details on this later on).

The lifetime contours in Fig.~\ref{fig:lifetimes} are shown for three cases when $N$ mixes dominantly with one of the $\nu_e$, $\nu_\mu$, and $\nu_\tau$ flavors according to Eq.~(\ref{nominal}) where $m_\nu \sim m_{\rm atm} \approx 0.05$ eV. The third case (mixing with $\nu_\tau$) results in a longer lifetime because of the phase space suppression of the $W$-mediated decay channel to $\tau$. The first two cases (mixing with $\nu_e$ and $\nu_\mu$) essentially result in the same lifetime as $M_N \gg m_\mu$.    

As seen in Fig.~\ref{fig:lifetimes}, boosted lifetimes of 1 to 10 seconds can be readily obtained for $M_N$ of few GeV and $M_{\rm DM} \sim 200-5000$ GeV. We, however, note that the rest frame lifetimes are in principle much shorter than 1 second. Since RH neutrinos can be produced with a significant abundance in the early universe (and may even reach thermal equilibrium), this ensures that that their decay does not pose any threat to big bang nucleosynthesis (BBN). 

\section{Signals from delayed decays}
\label{sect:spectra}

We now study the photon and neutrino signals from solar annihilation of DM into RH neutrinos within the DM mass range shown in Fig.~1\footnote{The photon signal from galactic DM annihilation and from dwarf spheroidals for heavier $N$ are discussed in~\cite{Excess,Rius}.}. In order to obtain the photon and neutrino spectra, we first implement the $\nu-N$ mixing in FeynRules~\cite{feynrules} to calculate the major decay modes of light $N$ and their corresponding branching fractions.
 
The decay is dominated by weak gauge interaction and the largest partial width is taken by virtual $W$ channel, $N \rightarrow l W^*$, where $W^*$ splits into either $\nu$ and a charged lepton, or a quark-antiquark pair. $N\rightarrow \nu Z^*$ has a smaller partial width as $Z^*$ due to larger $Z$ mass. We list all the significant modes in Table~\ref{tab:decay_channels}. Note that for a light RH neutrino mass, the partial decay widths of $N$ can be kinematically affected by the mass of final state particles. Also, since $N$ is a Majorana fermion, it decays into a given final state as well as its $CP$-conjugate final state at the same rate. 

\begin{table}[h]
\begin{tabular}{l|ccc}
\hline
Decay Mode & $\nu_e$ mixing case & $\nu_\mu$ mixing case & $\nu_\tau$ mixing case  \\
\hline
$N\rightarrow l (q\bar{q}')$ & 44.6\% & 44.3\% & 1.7\% \\
$N\rightarrow \nu_l (q\bar{q})$ & 19.9\% &  20.1\% &  53.5\% \\
$N\rightarrow l (\nu_{l'}\bar{l}')$ & 13.1\% &  13.2\% &  1.1\% \\
$N\rightarrow l \nu_{l}\bar{l}$ & 7.8\% &  7.6\% &  0\% \\
$N\rightarrow \nu_l (l'\bar{l}')$ & 1.6\% &  1.7\% &  8.6\% \\
$N\rightarrow \nu_l (\nu_{l'}\bar{\nu}_{l'})$ & 6.5\% &  6.6\% &  17.6\% \\
$N\rightarrow \nu_l \nu_l \bar{\nu}_l$ & 6.5\% &  6.6\% &  17.6\% \\
\hline
\end{tabular}
\caption{Decay channels of a 2.5 GeV RH neutrino via its mixing with the SM neutrinos. The partial widths are calculated at tree-level. Depending on the mass of $N$, not all channels are kinematically allowed. The final column differs because $N$ does not produce various final states that contain a $\tau$. The corrections from hadronization of $(q\bar{q})$ system in the hadronic modes are not included.}
\label{tab:decay_channels}
\end{table}

In the three-body decays of $N$, the SM neutrinos/leptons take a significant fraction of the total energy. The $\mu,~\tau$ leptons in the final state can further decay into neutrinos. Since $N$ has an energy $M_{\rm DM} \gg M_N$,
the charged leptons and neutrinos from $N$ decay acquire large energy due to the Lorentz boost. 
Neutrinos and photons produced from these energetic leptons (also from the hadronization and shower in semileptonic decays) yield the high energy neutrino and gamma ray signals for indirect searches.

The neutrino signal, due to IceCube detection thresholds and the fact that the neutrino scattering cross section increases with energy, receives most of the contribution from high energy part of the neutrino spectrum. In our case, the `hard' part of the spectrum is dominated by the neutrinos that directly emerge from the three-body decay of $N$. Secondary neutrino arise from the charged lepton and pion decays in the final state, but their contribution is subleading due to the much lower energy after several decay steps. We note that the hard part of the spectrum is not suppressed for delayed $N$ decays, unlike the standard scenario where neutrinos from DM annihilation are produced inside the Sun. For DM mass above few hundred GeV, the neutrino signal can be detected by IceCube.

The photon signal has two major components: (1) the charged lepton's bremsstrahlung radiation that yields a soft power-law shaped spectrum, and, (2) the neutral pion decay $\pi^0\rightarrow \gamma\gamma$ that arises abundantly in final states with a $\tau$ and also directly from the $N \rightarrow \nu Z^*$ decay channel. The bremsstrahlung contribution is mostly determined by a logarithmic dependence on the mass of the leading lepton in energy. As the leading lepton energy spectrum is universal (at least in the kinematically unsuppressed case), the bremsstrahlung strength can be directly inferred from the lepton flavor composition in $N-\nu$ mixings. In general, bremsstrahlung produces far fewer energetic photons than pion decay, but dominates the low energy part of the photon spectrum. This is different from angular-momentum restricted processes~\cite{bib:ib} where energetic internal bremsstrahlung becomes the leading contribution.

\begin{figure}[h]
\includegraphics[width=0.45\textwidth]{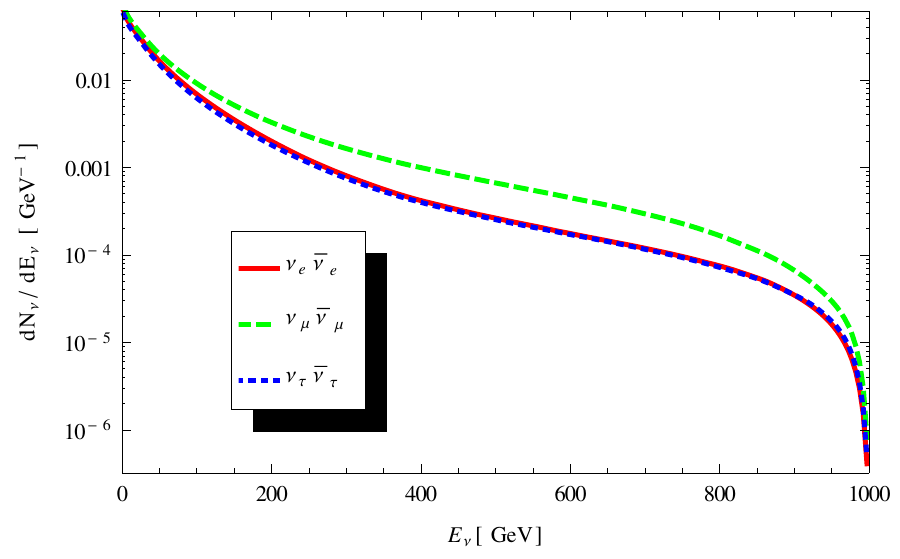}
\includegraphics[width=0.45\textwidth]{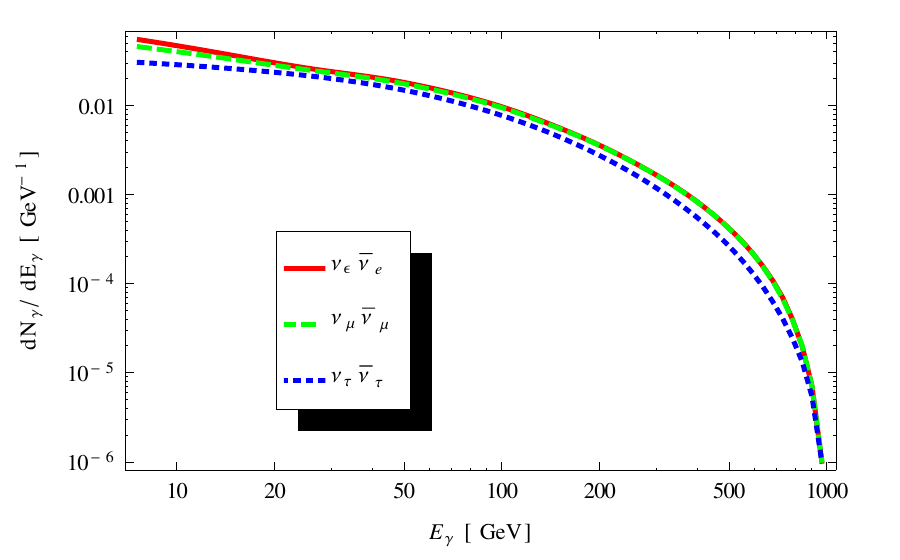}
\caption{
The spectra at the production point of $\nu_\mu$'s (top) and photons (bottom) from delayed decay of RH neutrinos produced in DM annihilation. In these plots, $M_{\rm DM} = 1$ TeV and $M_N = 2.5$ GeV.
The spectra are for $N$ mixing with each of the $\nu_e$, $\nu_\mu$, and $\nu_\tau$ flavors.}
\label{fig:prompt_spectra}
\end{figure}

The neutral pion contribution needs to be addressed more carefully. Apart from the usual energy fragmentation in $\tau$ decays that is well implemented in analysis tools, the GeV-scale mass of $N$ itself introduces decay channels like $N\rightarrow l/\nu +$~pions, where neutral pions can arises directly in the hadronic decays of $W^*$ and $Z^*$. 
Depending on the phase space in the hadronization process, the $q {\bar q}^{\prime}$ system preferably hadronizes into an angular momentum 1 final state. $\pi^0$ can either emerge from a multi-pion final state, or from the decays of excited spin-1 states like $\rho$, at non-trivial branching fractions. The $\left< \bar{q}\gamma^\mu q\right>$ form-factor of the semileptonic final state must be included and will yield a deviation in the partial decay width from a tree-level calculation (without hadronization), especially at $N$ masses below a few GeV, where QCD is non-perturbative. However, the hadronization modes vary with $N$ mass, and a thorough exploration of the hadronization corrections in $N$ decay would be beyond the scope of the current paper. In our numerical simulation of the prompt photon spectra, we carry out a tree-level decay calculation and rely on the string fragmentation in the PYTHIA8 package8~\cite{pythia8} for the hadronization process. Since the delayed decays of $N$ occur in vacuum, we require all unstable particles, in particular muons and mesons, to fully decay in the final state. The correction to the semileptonic decay branching ratio is not included. Noted that the pion multiplicity in semileptonic $N$ decays can affect the photon spectrum, we made a test at $m_N =m_\tau$, where $N$ decay is kinematically identical to $\tau$ decay, and found the pion multiplicity in semileptonic $N$ decays, in particular the sub-partition into single pion versus that in $\rho$ meson and multi-pion modes, agree very well with those in $\tau$ decay within few percent. Admittedly, for other and especially lower $N$ masses, the hadronic $N$ decay width and $\pi^0$ multiplicity may have non-negligible corrections. The results are shown in Fig.~\ref{fig:prompt_spectra} where the neutrino and gamma ray spectra derive from the delayed decay of $N$ from the annihilation of 1 TeV DM particles. We picked a benchmark point at $M_N = 2.5$ GeV, above the $\tau$ mass so that the RH $N$ decay into $\tau$ is kinematically allowed for $\nu_\tau$-$N$ mixing cases.


\begin{figure}[h]
\includegraphics[width=0.4\textwidth]{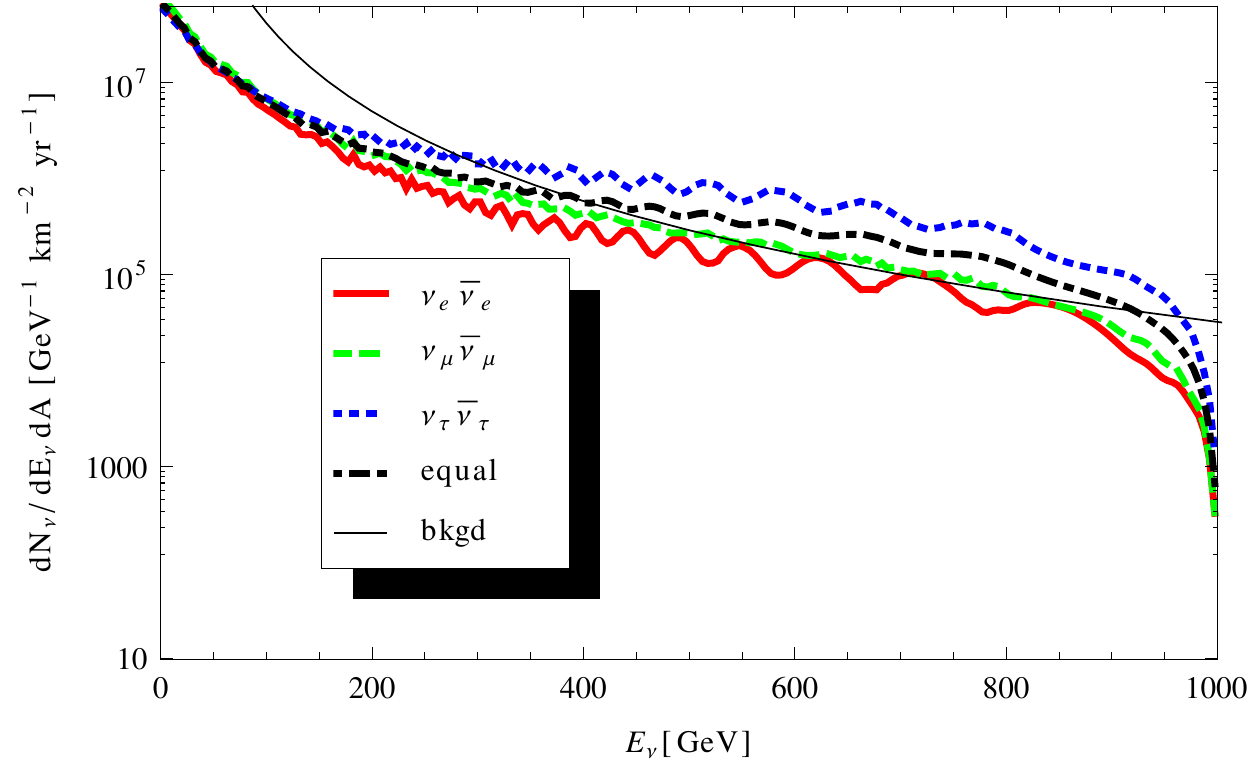}
\includegraphics[width=0.4\textwidth]{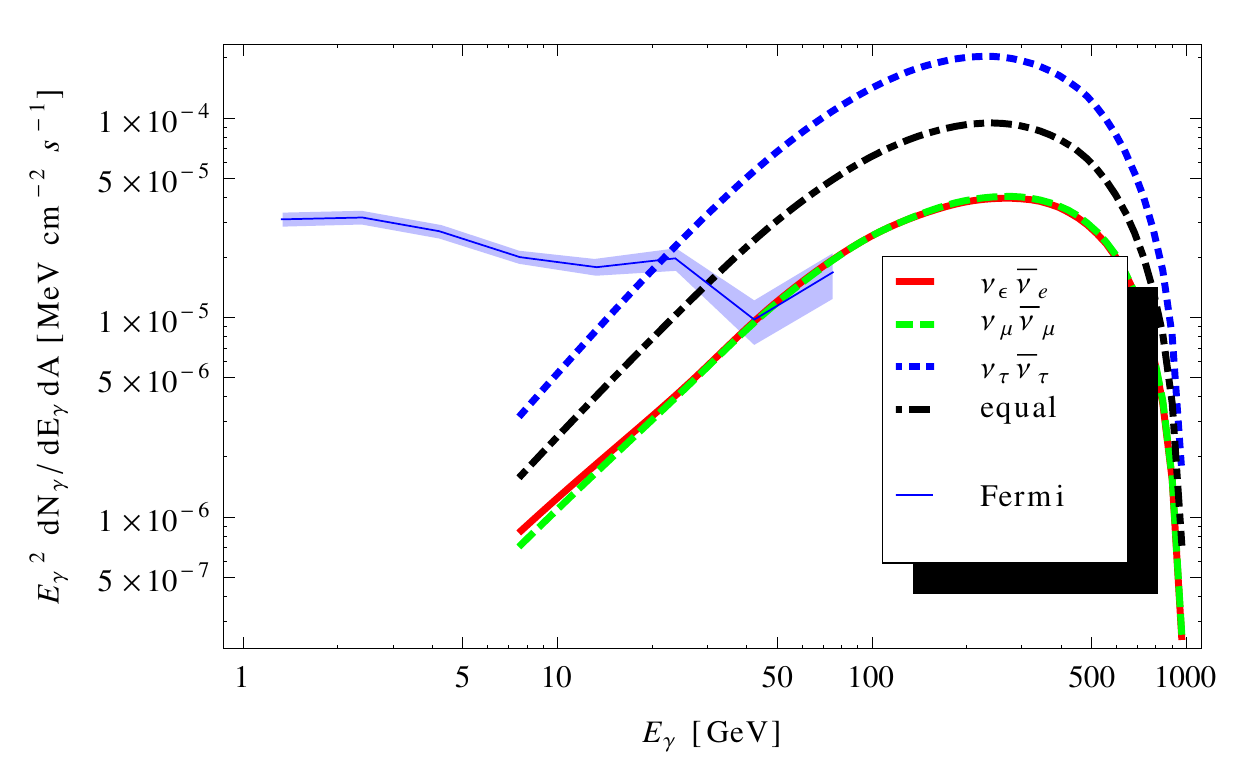}
\caption{The spectra at the detection point of $\nu_\mu$'s (top) and photons (bottom) from delayed decay of RH neutrinos produced in DM annihilation. In these plots, $M_{\rm DM} =1$ TeV and $M_N = 2.5$ GeV.
For normalization, we assume a total annihilation inside the Sun to be $1.5\times 10^{19}$ per second. The ``equal'' curves is for the case when $N$ mixing with all of the $n_e$, $\nu_\mu$ and $\nu_\tau$ flavors equally. The neutrino background is atmospheric neutrinos~\cite{atmosphericBackground}. The shaded region around the Fermi data shows the uncertainty from Ref.~\cite{fermiSun}. 
}
\label{fig:spectra}
\end{figure}

\begin{figure}[h]
\includegraphics[width=0.4\textwidth]{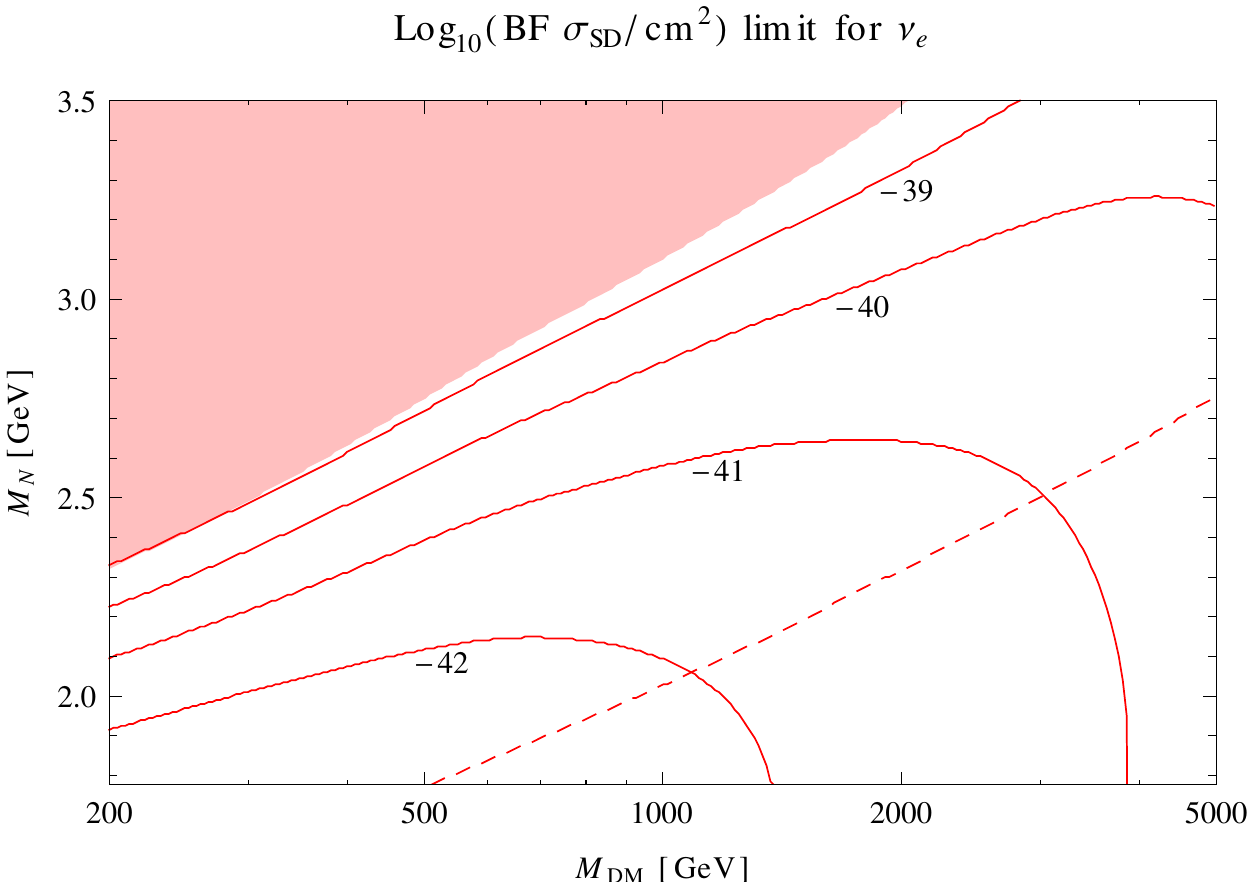}
\includegraphics[width=0.4\textwidth]{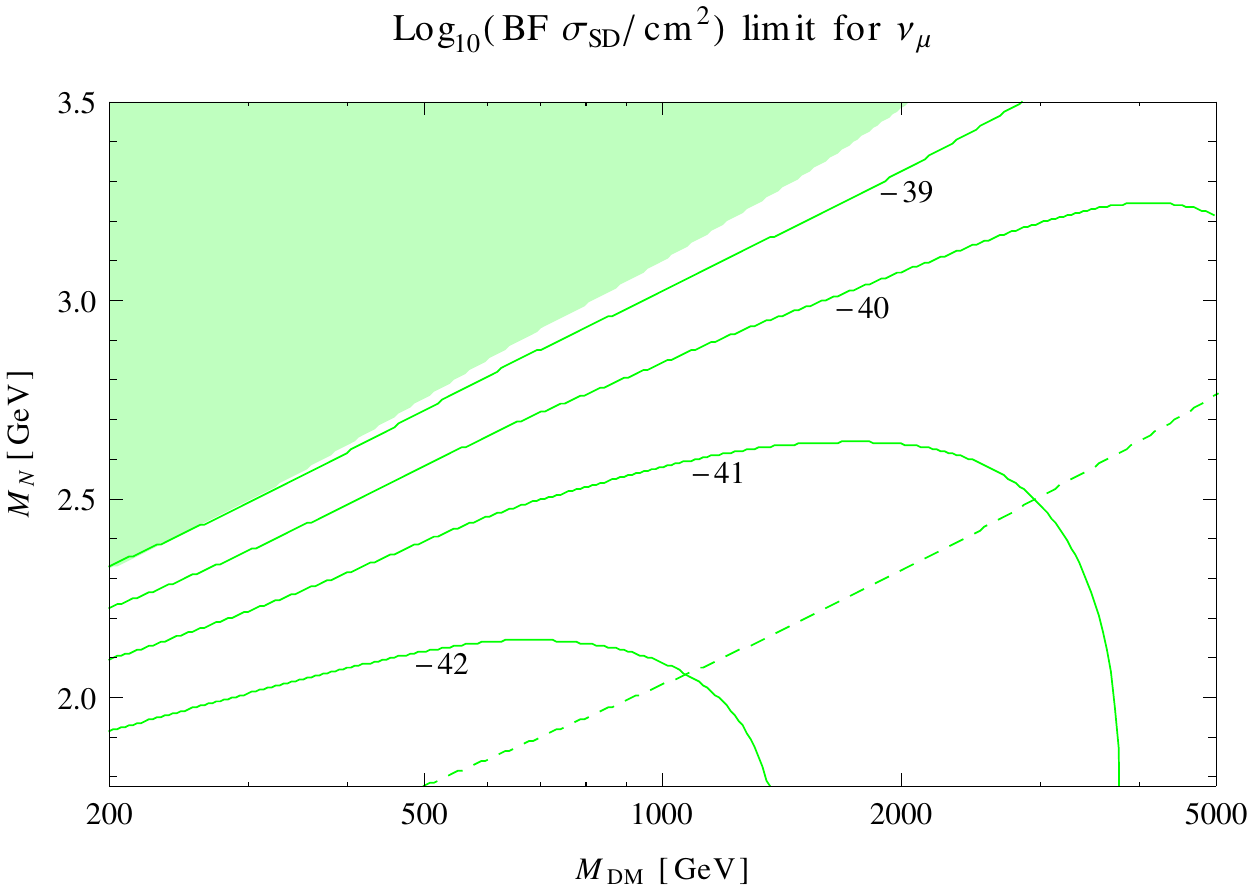}
\includegraphics[width=0.4\textwidth]{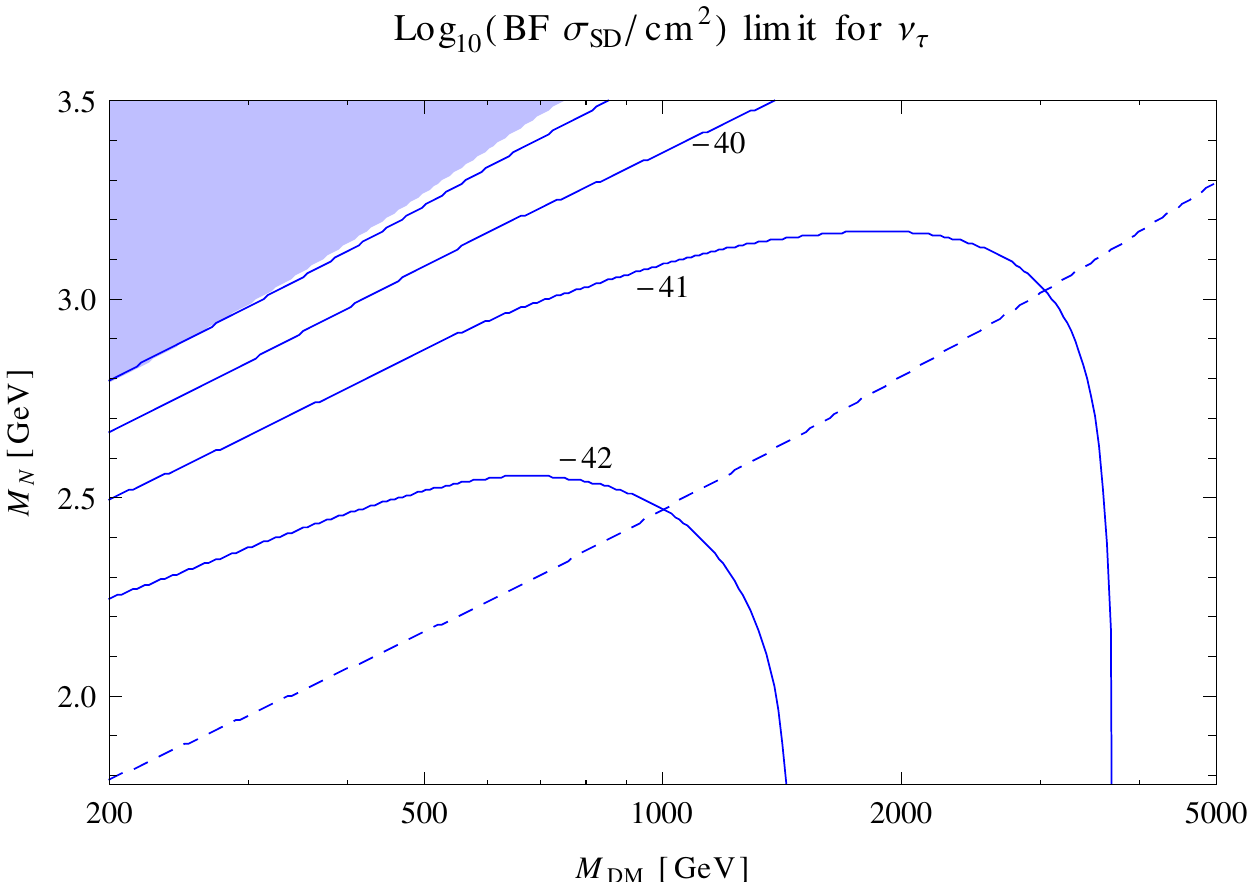}
\caption{The $90\%$ C.L. contours for $\sigma_{\rm SD}$ using the Fermi-LAT limits on the photon signal from solar DM annihilation to light RH neutrinos.
The shaded regions represent the parameter space where Fermi-LAT constraint on $\sigma_{\rm SD}$ is less stringent than that from PICO-60 limits~\cite{pico}. 
The boosted $N$ decay length is inside the photosphere above the dashed line. BF denotes the dark matter annihilation branching fraction into $N$ pairs.}
\label{fig:photonSD}
\end{figure}

\begin{figure}[h]
\includegraphics[width=0.4\textwidth]{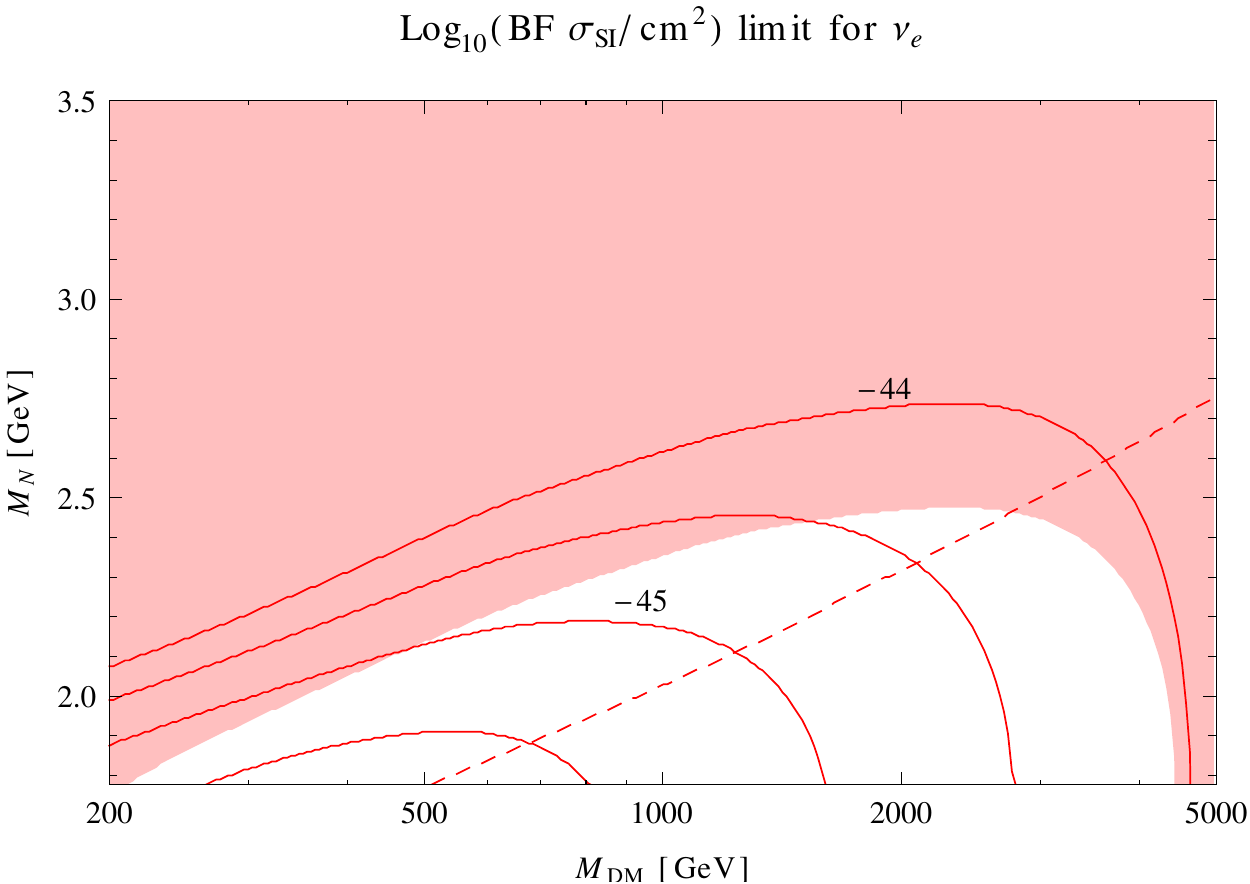}
\includegraphics[width=0.4\textwidth]{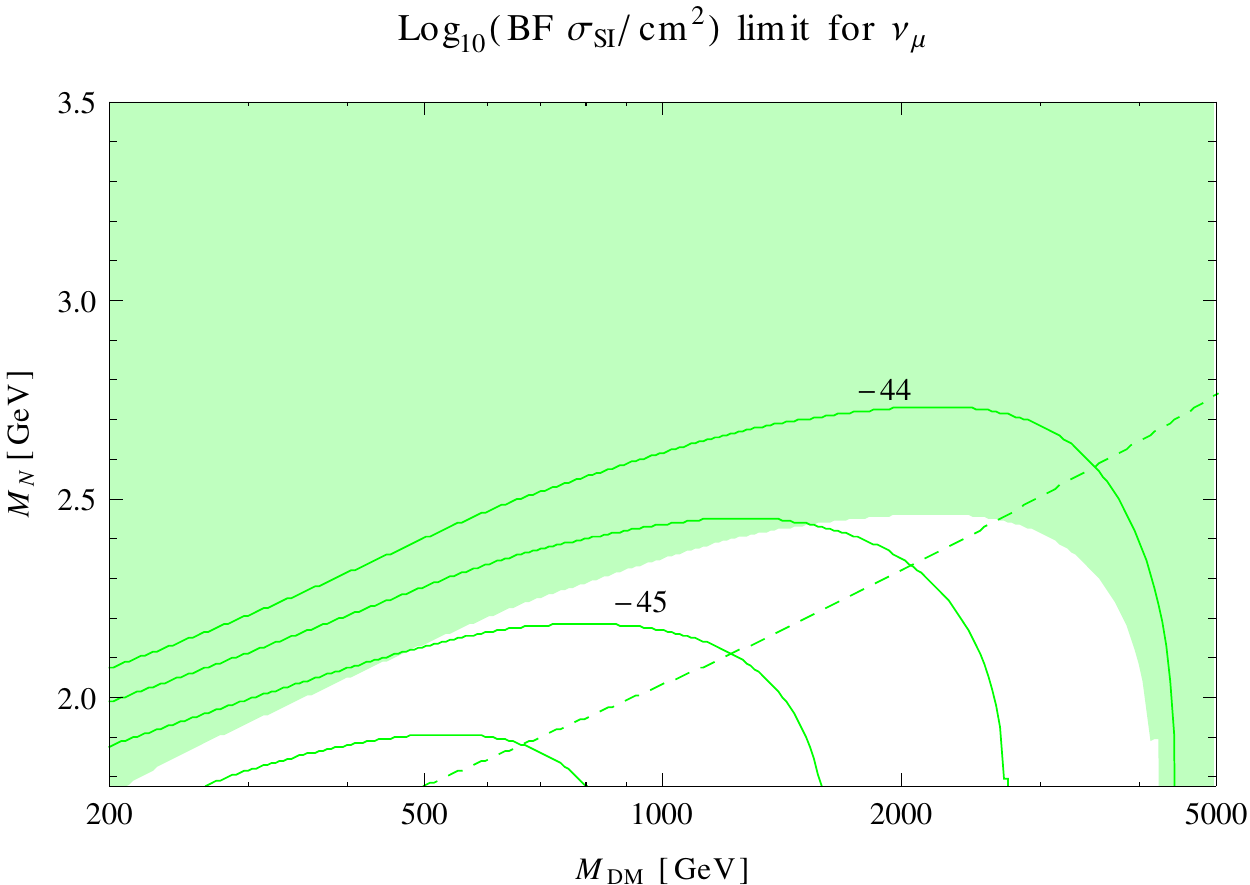}
\includegraphics[width=0.4\textwidth]{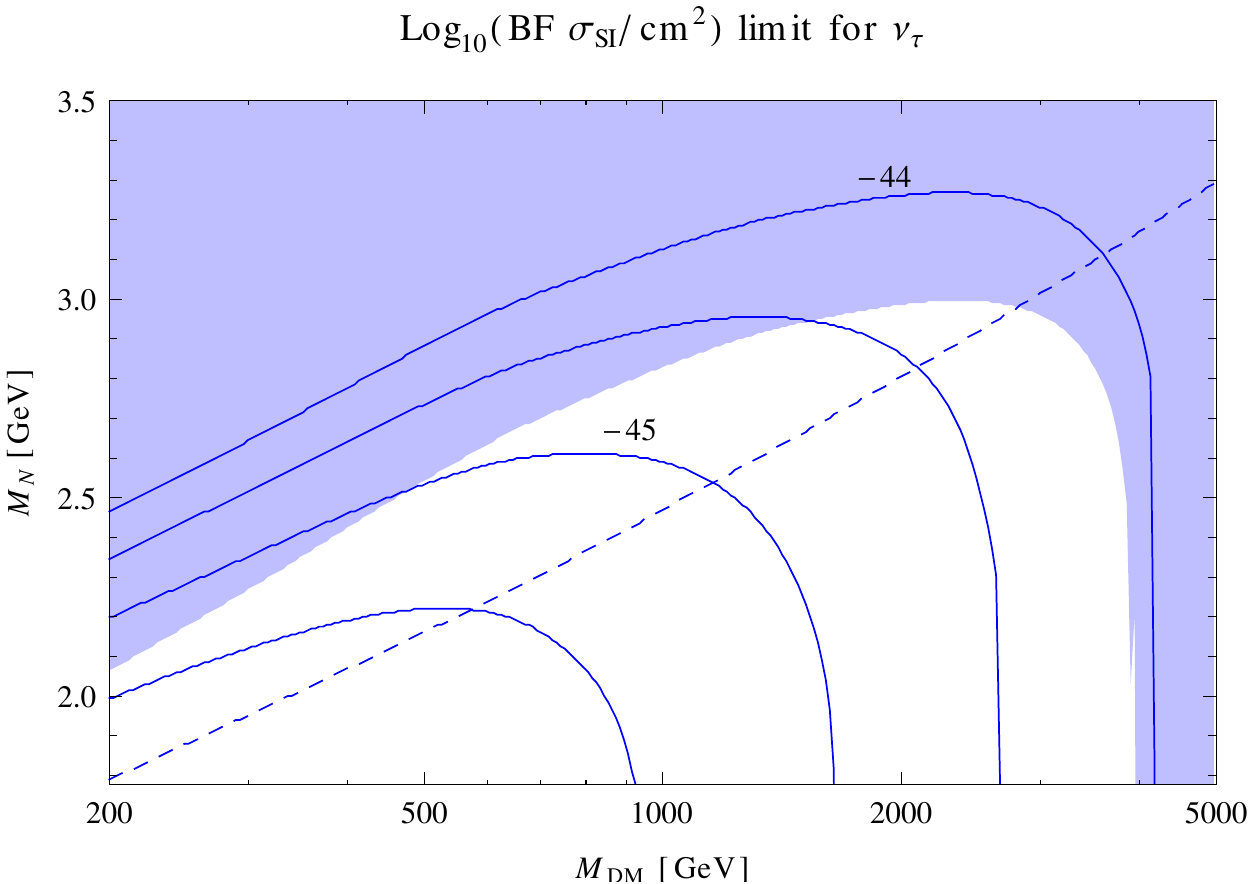}
\caption{Same as Fig.~\ref{fig:photonSD} but for $\sigma_\mathrm{SI}$. The shaded regions represent the parameter space where Fermi-LAT constraint on $\sigma_{\rm SI}$ is less stringent than that from LUX limits~\cite{LUX}.}
\label{fig:photonSI}
\end{figure}

\begin{figure}[h]
\includegraphics[width=0.4\textwidth]{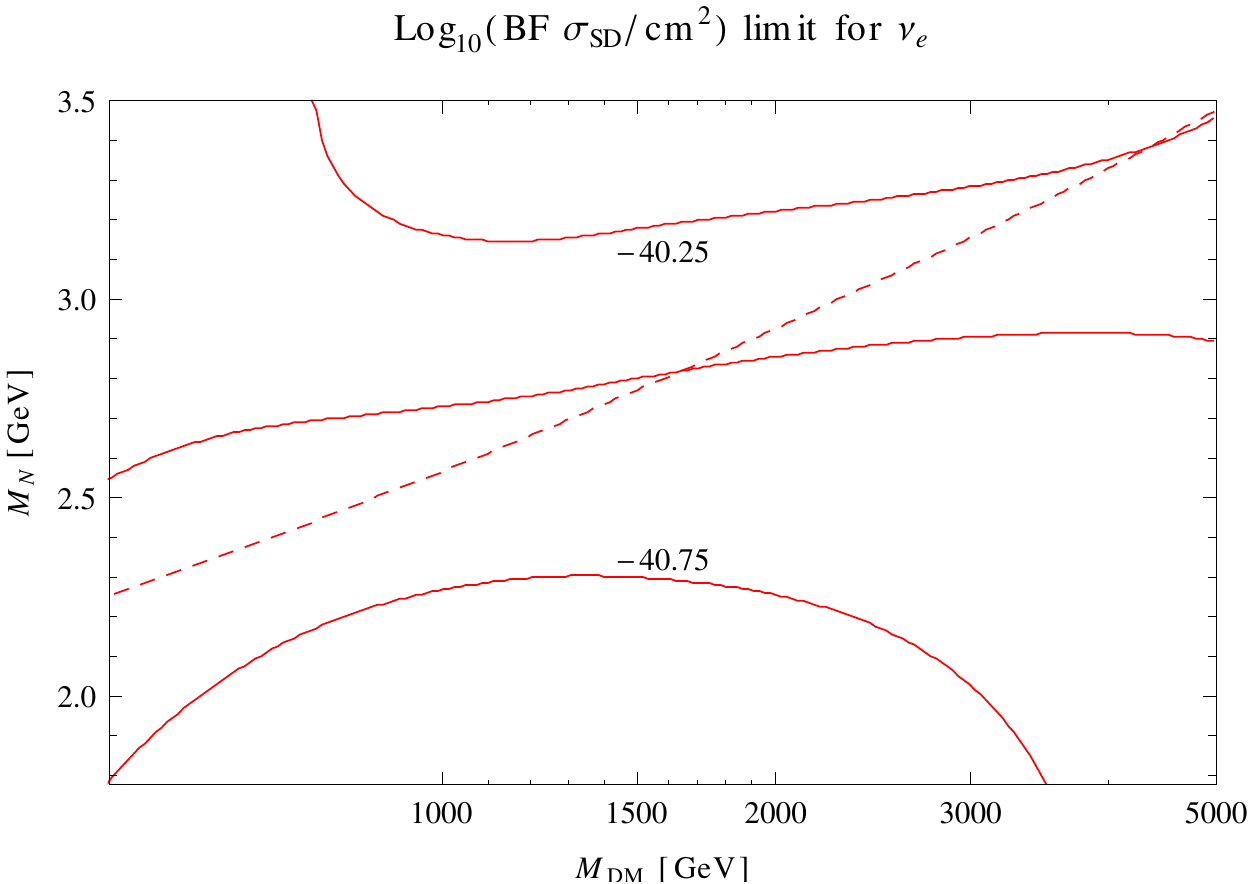}
\includegraphics[width=0.4\textwidth]{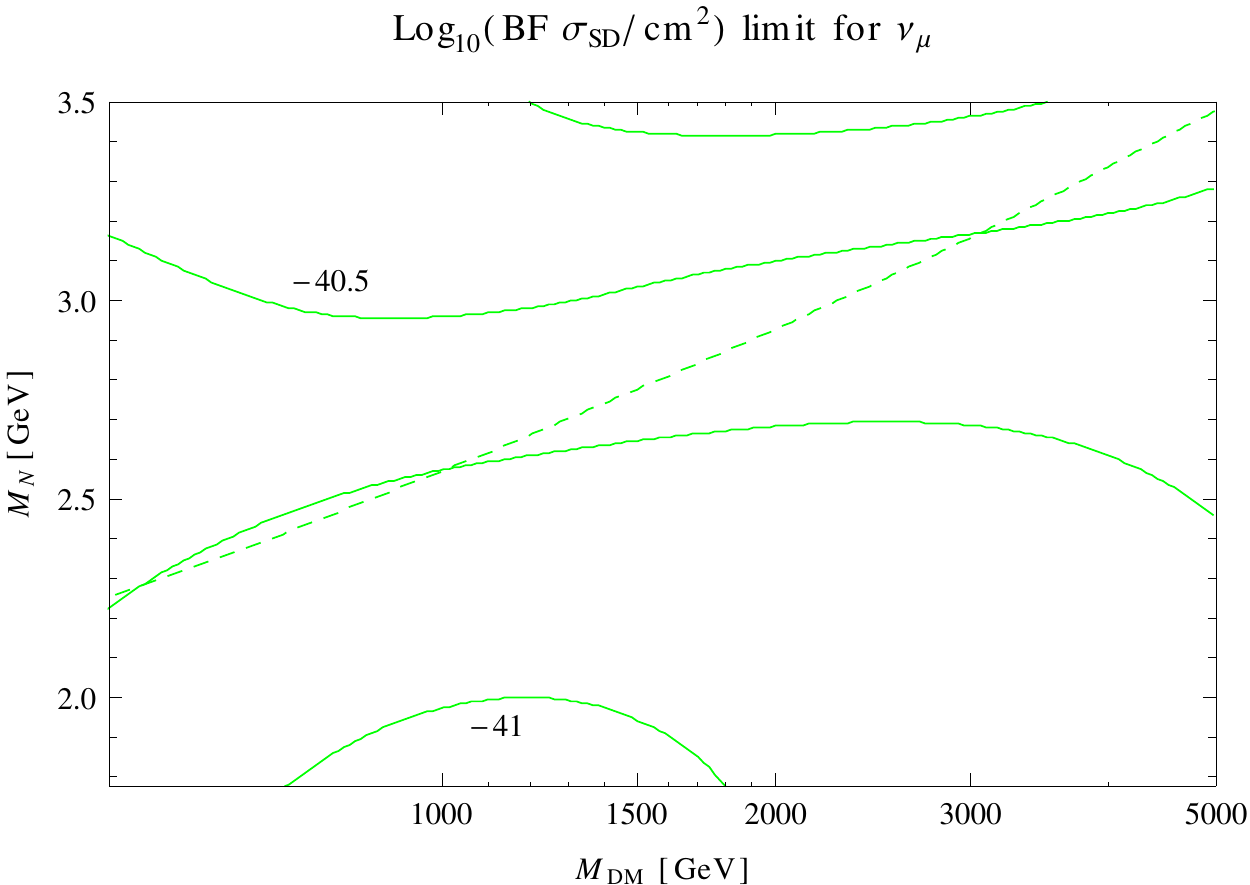}
\includegraphics[width=0.4\textwidth]{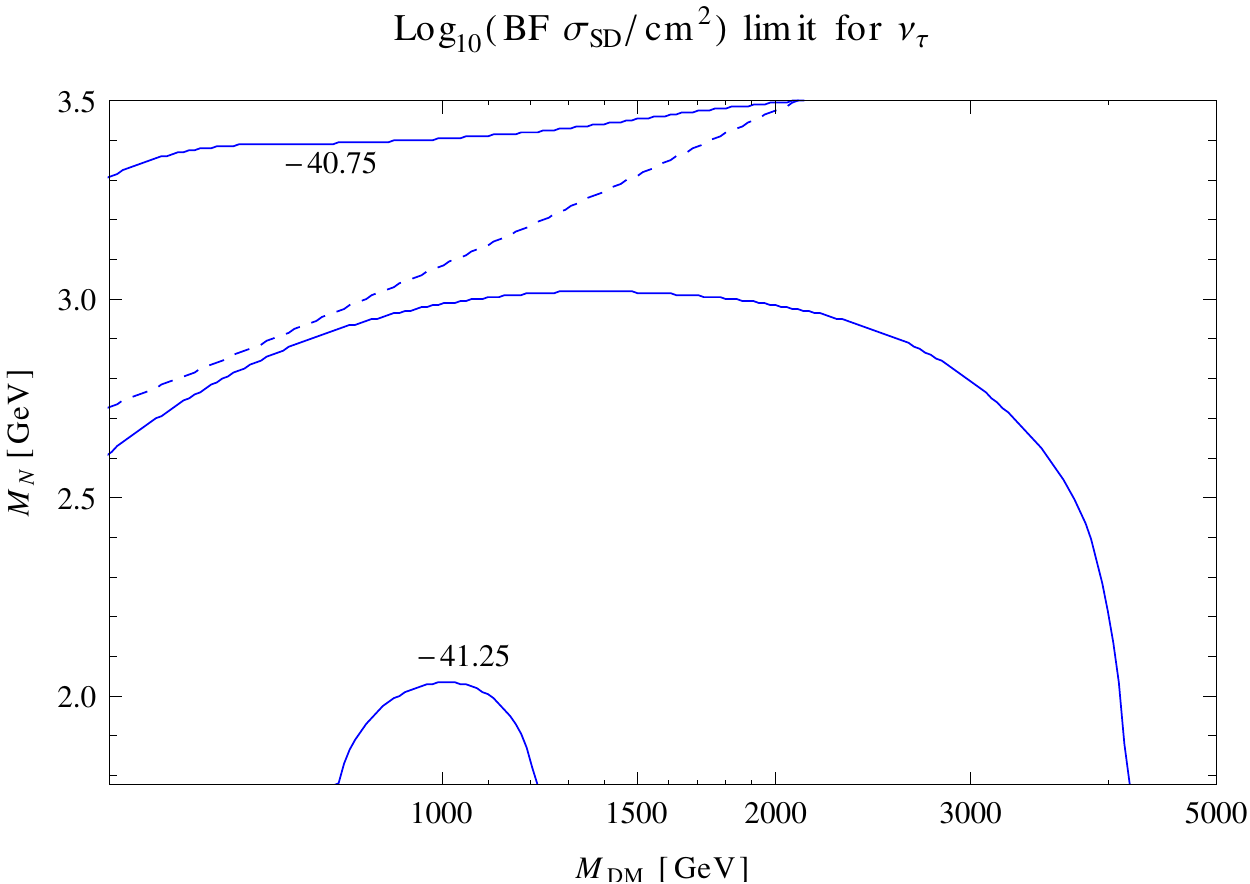}
\caption{The $90\%$ C.L. contours for $\sigma_{\rm SD}$ using IceCube limits on the neutrino signal from solar DM annihilation to light RH neutrinos.
The dashed line now marks when the characteristic length of the decays is at the 200,000 km cut-off.}
\label{fig:neutrinoSD}
\end{figure}

\begin{figure}[h]
\includegraphics[width=0.4\textwidth]{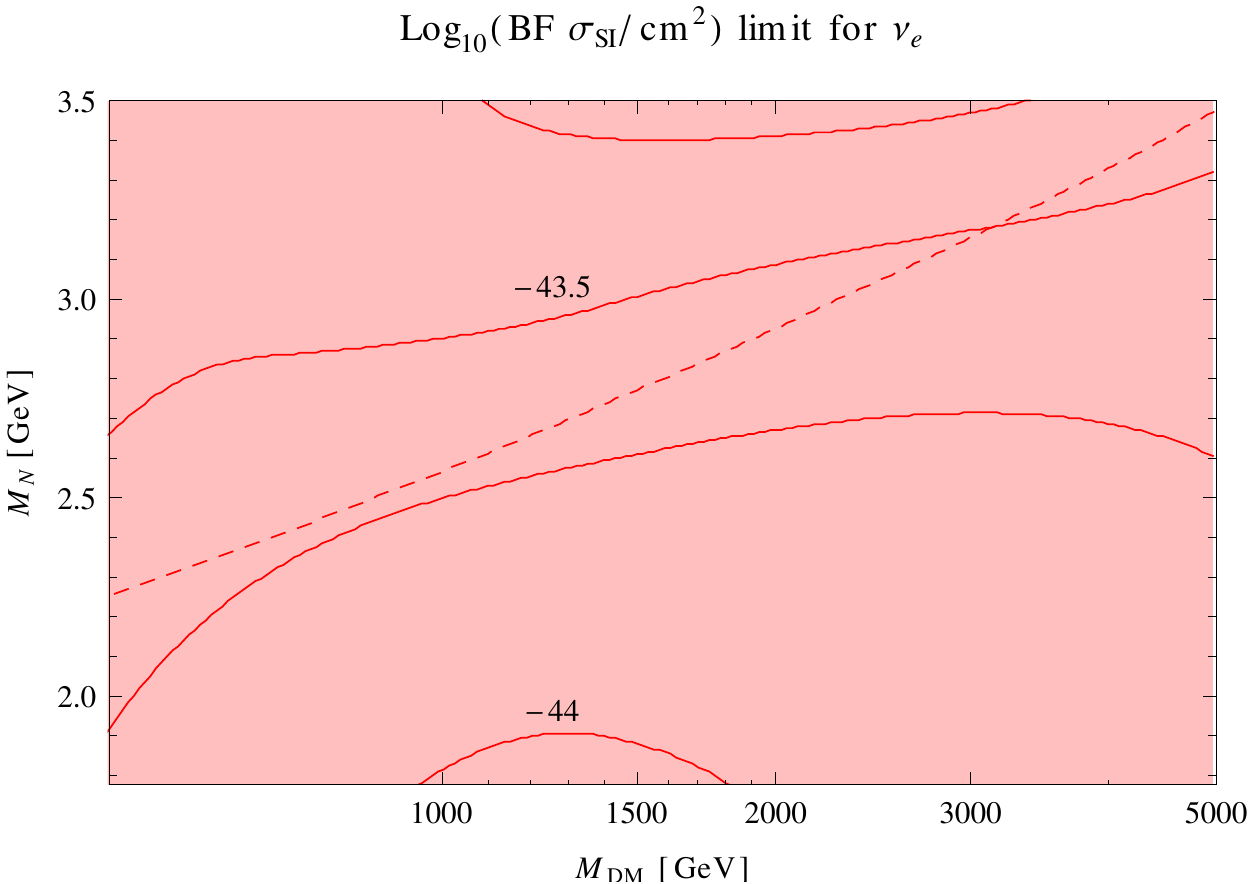}
\includegraphics[width=0.4\textwidth]{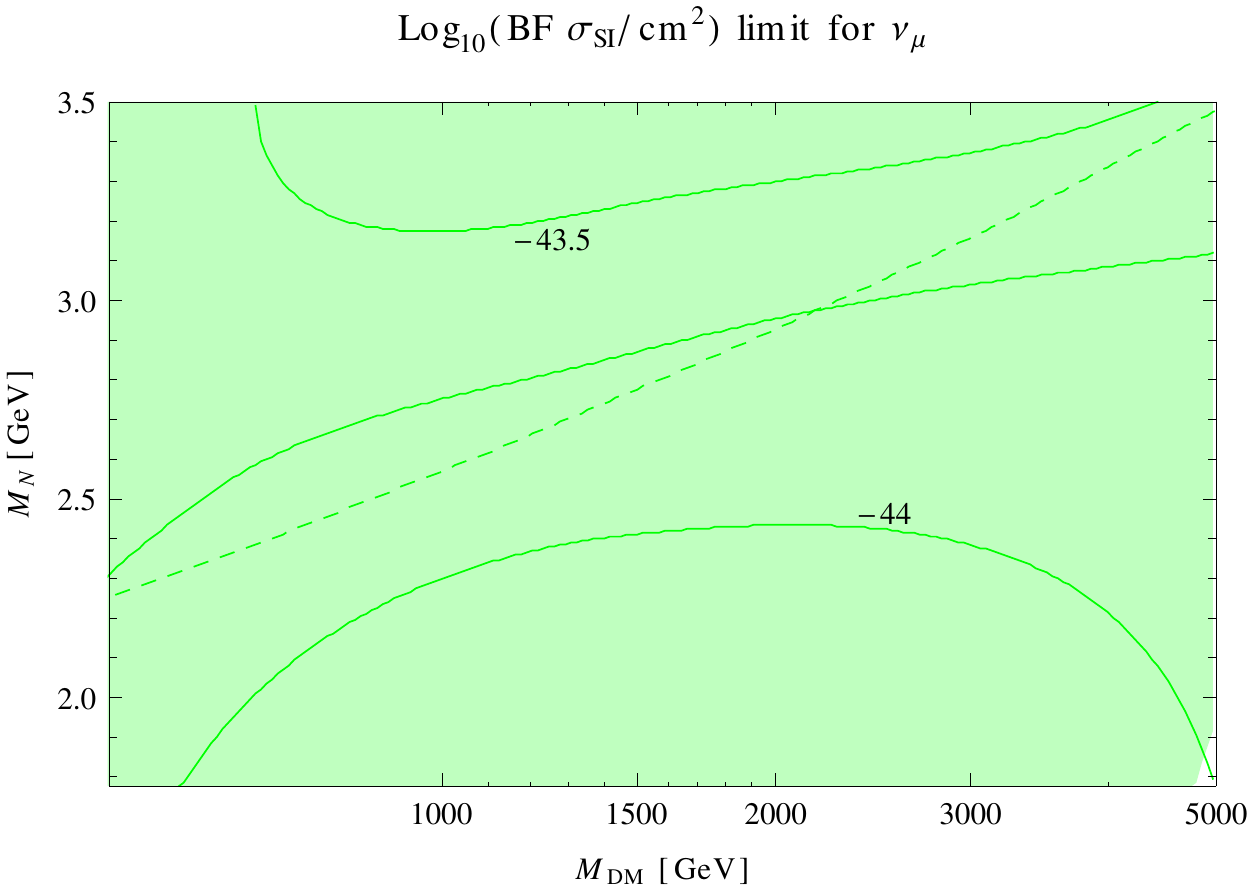}
\includegraphics[width=0.4\textwidth]{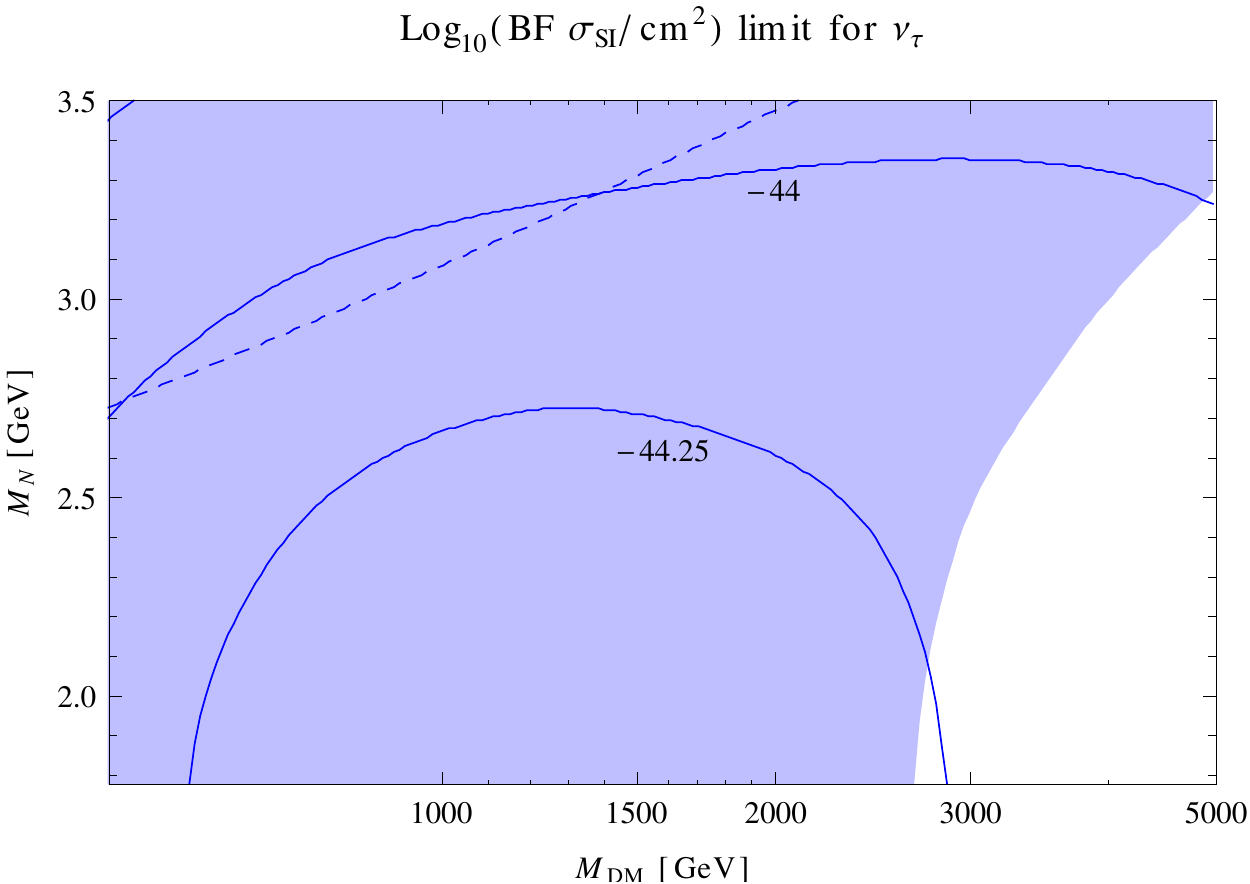}
\caption{Same as Fig.~\ref{fig:neutrinoSD} but for $\sigma_{\rm SI}$. The shaded regions represent the parameter space where IceCube constraint on $\sigma_{\rm SI}$ is less stringent than that from LUX limits~\cite{LUX}.  
}
\label{fig:neutrinoSI}
\end{figure}

In the calculation of the signal flux, we make the assumption that 
the DM capture and annihilation inside the Sun have reached an equilibrium. For a given $\sigma_\mathrm{SD}$ or $\sigma_\mathrm{SI}$, we use DarkSUSY~\cite{darksusy} to calculate the minimum annihilation rate $\langle \sigma v \rangle$ for equilibrium. Fig.~\ref{fig:spectra} illustrates sample spectra that correspond to $\sigma_\mathrm{SI}=8.5\times 10^{-45}\ \mathrm{cm}^2$ for $M_{\rm DM} = 1$ TeV, which saturates the latest LUX bound~\cite{LUX}.

As the $N$ decay occurs between the solar center and the Earth, and the source intensity decreases exponentially over distance. The large Lorentz boost along the line-of-sight forces most of the $N$ decay products to fall along the forward direction, pointing away from the Sun. Very soft photons may still deviate away from the line-of-sight direction. We adopt an half-cone radius cut of 1.5$^\circ$ as given in Fermi-LAT observation on solar disk~\cite{fermiSun}, and integrate over the decay's source intensity within. We have checked this 1.5$^\circ$ angular cut sufficiently covers the angular spread for the photon energy in Fermi-LAT's observation range (0.2-200 GeV).

For short boosted lifetime, a fraction of decays may still happen inside the Sun. For gamma rays  we take a simple cut at the Sun's photosphere radius $R_{\odot} \approx 700,000$ km, and consider only the decays outside the photosphere contribute to the signal. However, the situation is more complicated for the neutrino signal. Since attenuation of the neutrino flux is energy dependent, the distortion in the energy spectral shape depends on the DM mass, the prompt spectral shape, and the distance inside the solar medium. Due to flavor oscillations, the attenuation effects also differ between different neutrino flavors. Since our focus is on long-lived $N$ that decays mostly outside the Sun, we make an approximate cuts-off at $R \approx 200,000$ km and only include neutrino sources outside this radius, where the neutrino fluxes are considered unattenuated. This selection is based on the fact that at TeV energies and below, the flux of neutrinos produced above this distance is suppressed by less than 10$\%$ in the solar medium. This approximation is a reasonable simplification yet it may underestimate the neutrino signal and leads a more conservative constraint for intermediate $N$ decay lengths close to the $R \approx 200,000$ km cut-off.

We the calculate the signal spectra for all three flavor-mixing scenarios in $\nu-N$ for DM masses between 10 GeV and 5 TeV. Vacuum oscillation is included for neutrino propagation to the Earth, where we used SM neutrino parameters from the latest global fit~\cite{nuData}. See Appendix~\ref{app:neutrino} for details. As IceCube has angular sensitivity only for muon track-events from the charged-current interaction of $\nu_\mu$', the different flavor schemes in $N-\nu$ mixing can yield quite different $\nu_\mu$ fluxes both at the source and at the Earth. 

A few comments are due on some features in Figs.~\ref{fig:prompt_spectra},~\ref{fig:spectra}. First, contrary to what one might expect, it is seen in Fig.~\ref{fig:prompt_spectra} that the $N-\nu_\tau$ mixing produces less photons than the $N-\nu_\mu$ and $N-\nu_e$ mixing cases. The reason is that, as mentioned above, neutral pions that are the dominant source of photon production mainly arise from $N \rightarrow l  W^*$ decays. However, the $N \rightarrow \tau W^*$ channel is kinematically suppressed relative to $N \rightarrow \mu W^*$ and $N \rightarrow e W^*$ channels due to the larger value of $m_\tau$. On the other hand, this kinematic suppression also results in a smaller total decay rate for $N$ in the case $N$ mainly mixes with $\nu_\tau$. The longer lifetime of $N$ in this case implies a larger number of decays outside the Sun, and hence a stronger photon signal as seen in Fig.~\ref{fig:spectra}. Second, we observe an oscillatory pattern at energies above few hundred GeV in the neutrino spectra at energies above few hundred GeV at the detection point in Fig.~\ref{fig:spectra}. This is because at such energies neutrino vacuum oscillations due to the solar mass splitting are not averaged out over the $\sim 3,000,000$ km variation in the Earth-Sun distance within half a year (which the IceCube uses to collect data in the direction of the Sun).

\section{Experimental constraints}
\label{sect:indirect}

Assuming an equilibrium between DM capture and annihilation in the Sun, and using the signal spectra calculations in the previous section, we derive constraints on $\sigma_{\rm SI}$ and/or $\sigma_{\rm SD}$ from Fermi-LAT and IceCube limits on the photon and neutrino signals respectively. For a DM mass range of 0.2-5 TeV, the Fermi-LAT constraints on photonic signals are shown in Figs.~\ref{fig:photonSD} and ~\ref{fig:photonSI}, and IceCube's constraint on signal neutrino fluxes are shown in Figs. ~\ref{fig:neutrinoSD} and ~\ref{fig:neutrinoSI}. In each figure, the three panels illustrate exclusive $N-\nu_e$ (upper), $N-\nu_\mu$ (middle) and $N-\nu_\tau$ (lower) mixing cases separately, where in each case $N$ only mixes with one SM neutrino flavor. A nominal annihilation cross section $\langle \sigma v \rangle = 3 \times 10^{-26}$ cm$^3$ s$^{-1}$ can achieve the capture-annihilation equilibrium for the scattering cross sections $\sigma_{\rm SI}$ and $\sigma_{\rm SD}$ in these results. Details on the photon and neutrino signal flux calculation are discussed in Appendix~\ref{app:photon} and ~\ref{app:neutrino}.

The contours for $\sigma_\mathrm{SD}$ in Fig.~\ref{fig:photonSD} show the spin-dependent scattering cross sections ruled out at 90\% C.L. by Fermi-LAT, where the constraint is dominated by the highest energy bin in the solar gamma ray data. Details of fitting is discussed in Appendix~\ref{app:photon}. Here we can make a comparison with the direct detection limit: at each DM mass and its direct detection constraint on scattering cross-section, a $m_N$ mass can be obtained where Fermi-LAT constrains to the same cross-section. Below this $m_N$ the Fermi-LAT constraint is stronger than that from direct detection (at the same DM mass). This is shown by the boundary of the shaded regions, that non-shaded region represents a better constraint on the $\sigma_{\text{SD}}$ in comparison to direct detection limits from PICO-60~\cite{pico}. For spin-independent $\sigma_{\text{SI}}$, a similar comparison can be done with LUX~\cite{LUX} results, as illustrated in Fig.~\ref{fig:photonSI}.

We see that the 90\% C.L. values in Figs.~\ref{fig:photonSD},~\ref{fig:photonSI} become tighter when $M_N$ decreases for a given $M_{\rm DM}$. This is because of increased $N$ lifetime $\tau_N$, that results in a larger fraction of RH neutrinos decaying outside the photosphere. Increasing $M_{\rm DM}$, while keeping $M_N$ constant, initially leads to stronger bounds as a larger Lorentz boost $\gamma= M_{DM}/M_N$ also leads to longer decay length for $N$. However,  contours reverse and quickly drop with further increase in $M_{\rm DM}$. This is because (1) at very large DM masses the peak flux in photons becomes too energetic and moves above Fermi-LAT's energy reach; (2) the solar DM capture rate also drops at large DM mass. We note that the tightest bounds are obtained in the case that $N$ mixes with $\nu_\tau$ as the $\tau$ decay also yields abundant energetic photons.    

To constrain the neutrino signal, we use the 90\% C.L. IceCube bound in Ref. and compare the muon event rates per volume per year for detection. 
The bound on $\sigma_{SI},\sigma_{SD}$ assuming DM DM$\rightarrow NN$ annihilation channel can be obtained by scaling these scattering cross sections so that they yield the same muon event rates as the channels given in Ref.~\cite{ICBounds}. We have scaled $\sigma_\mathrm{SD}$ and $\sigma_\mathrm{SI}$ to obtain the same number of muons (plus antimuons) in our case. Detail of the analysis is discussed in Appendix ~\ref{app:neutrino}.  The constraint contours for $\sigma_\mathrm{SD}$ (Fig.~\ref{fig:neutrinoSD}) and $\sigma_\mathrm{SI}$ (Fig.~\ref{fig:neutrinoSI}) show the value of respective cross sections that are ruled out at 90\% C.L. according to IceCube's sensitivity. As we see in Fig.~\ref{fig:neutrinoSD}, IceCube does much better than direct detection experiments in the case of $\sigma_{\rm SD}$ as PICO-60 limits are much weaker. The situation is different for $\sigma_{\rm SI}$, see Fig.~\ref{fig:neutrinoSI}, where the shaded regions are already ruled out by the LUX results~\cite{LUX}.

Similar to the photon case, the 90\% C.L. values in Figs.~\ref{fig:neutrinoSD},~\ref{fig:neutrinoSI} get tighter as $M_N$ decreases for a given $M_{\rm DM}$ because it leads to a longer $\tau_N$ resulting in a larger fraction of RH neutrinos decaying outside the 200,000 km cut-off. However, the contours above and below the dashed line (denoting the lifetime corresponding to a characteristic radius of 200,000 km) behave differently with increasing $M_{\rm DM}$ when $M_N$ is kept constant. For those above the line, a significant fraction of $N$ decays happen inside this radius. Therefore, the bound initially gets weaker as a larger Lorentz boost in the energy of neutrinos results in more interactions with the solar medium that suppresses the neutrino signal. This reverses for larger values of $M_{\rm DM}$ for which a larger fraction of RH neutrinos decay outside the 200,000 km radius. The situation is opposite for the contours below the dashed line, for which the majority $N$ decays occur outside this radius. Increasing $M_{\rm DM}$ initially results in more energetic neutrinos, and hence a stronger neutrino signal at IceCube. Further increase in $M_{\rm DM}$, however, leads to a smaller neutrino flux as a smaller number of DM particles are captured. As in Figs.~\ref{fig:photonSD},~\ref{fig:photonSI}, the tightest bounds arise in the case that $N$ mainly mixes with $\nu_\tau$ because of additional neutrinos from $\tau$ decay. Interestingly, as shown by the non-shaded region in Fig.~\ref{fig:neutrinoSD}, IceCube can be more stringent than LUX in constraining $\sigma_{\rm SI}$ in this mixing scenario.   

We see in Figs.~\ref{fig:photonSD},~\ref{fig:neutrinoSD} that the 90$\%$ C.L. values from the neutrino signal on $\sigma_{\rm SD}$ are stronger than the direct detection bounds in the entire parameter space considered here. In the case of $\sigma_{\rm SI}$, the photon signal sets tighter limits than direct detection experiments in a sizable region of the parameter space for DM masses up to $\sim 4000$ GeV. The photon and neutrino signals combined together do better than direct detection experiments in more than half of the parameter space for the entire DM mass range $200-5000$ GeV in the case of $N-\nu_\tau$ mixing. The constraints from the photon signal can be extended to even higher DM masses by future data from gamma ray observatories like HAWC~\cite{HAWC} and DAMPE~\cite{Feng:2014vza}, which can detect photons with higher energies than those detectable by Fermi-LAT.

\section{Conclusions}
\label{sect:conclusion}

In this paper, we have performed a study of indirect detection signals from solar annihilation of DM into RH neutrinos $N$ with a mass $M_N\sim$1-5 GeV. These RH neutrinos dominantly decay via off-shell $W$ and $Z$ due to their small mixing with the LH neutrinos. For DM mass $M_{\rm DM}$ from 200 GeV to 5 TeV, and nominal value of mixing expected in Type-I seesaw, the RH neutrinos can have a lifetime $\tau_N \sim$1-10 s and escape the Sun before decaying. The delayed decays then give rise to a photon signal in the direction of the Sun, as well a neutrino signal that is not attenuated by absorption and scattering in the solar medium.

The strongest signals are obtained in the case that RH neutrinos produced from DM annihilation mainly mix with $\nu_\tau$. Then, for $M_N > m_\tau$, delayed decays of $N$ produce taus whose decay produces more photons (due to their semileptonic decays) and neutrinos than muons and electrons. We have used the Fermi-LAT and IceCube limits on the photon and neutrino signals, respectively, in the direction of the Sun to constrain the product of the branching fraction of DM annihilation to RH neutrinos and the DM-nucleon elastic scattering cross sections at 90$\%$ C.L.

The Fermi-LAT sets stringent bounds on both $\sigma_{\rm SI}$ and $\sigma_{\rm SD}$. It gives rise to significantly tighter limits on $\sigma_{\rm SI}$ than the most stringent ones from direct detection experiments~\cite{LUX,PandaX} for DM masses from $\sim 200$ GeV up to 4 TeV. The IceCube also sets limits on $\sigma_{\rm SI}$ that are stronger than those in~\cite{LUX,PandaX} for DM masses above 4 TeV, in the case that $N$ mixes mainly with $\nu_\tau$. Both Fermi-LAT and IceCube set bounds on $\sigma_{\rm SD}$ that are much tighter than the strongest limits from direct detection experiments~\cite{pico}.

The neutrino signal from delayed decays of light RH neutrinos can probe the DM-nucleon elastic scattering cross sections for DM masses up to several TeV. This is much better than the usual scenario where neutrinos produced from solar DM annihilation are highly suppressed because of absorption and scattering in the Sun. The photon signal can also lead to stronger constraints at larger values of $M_{\rm DM}$ by using data from experiments like HAWC and DAMPE that are sensitive to gamma rays with higher energies than those detectable by Fermi-LAT.

\medskip
{\bf Acknowledgements}

The work of R.A. and B.K. is supported in part by NSF Grant No. PHY-1417510. Y.G. thanks the Mitchell Institute for Fundamental Physics and Astronomy (MIFPA), and Wayne State University for support. The work of S.S. is supported by the Los Alamos National Laboratory LDRD Program.

\appendix

\section{Photon signal flux}
\label{app:photon}

The total rate of annihilation events in the Sun is given~\cite{JKG} as,
\be
\Gamma_{\rm ann} = {C \over 2} {\rm tanh}^2(t/\tau) .
\ee
where $C$ is the capture rate of DM particles by the Sun. Here $\tau$ denote a time scale the equilibrium is established. $C$ depends on the scattering cross sections $\sigma_{\rm SI}$, $\sigma_{\rm SD}$, and is given in Ref.~\cite{JKG},
\be
C=\left\{
\begin{array}{l}
4.8\times10^{28}s^{-1} \frac{\rho_{0.3}}{\bar{v}_{270} m_{\chi^0}} \sum_{i} F_i f_i \phi_i  \frac{\sigma^{SI}_i }{m_{N_i}}
S\left({\scriptsize \frac{m_{\chi^0}}{m_{N_i}}}\right)\,,\\ 
1.3\times10^{29}s^{-1}\frac{\rho_{0.3}}{\bar{v}_{270} m_{\chi^0}}\sigma^{SD}_{H} S\left(  \frac{m_{DM}}{m_{N_i}}\right)\,,
\end{array}
\right.
\label{eq:ann}
\ee
for spin-independent and spin-depend scattering contributions respectively. The subscript index $i$ sums over the nuclear elements in the Sun. Here $\{N_i\}$ denotes the nucleus of $i$th element, not to be confused with the RH neutrino.  $\rho_{0.3}$ is the local DM halo density in units of 0.3 GeV/cm$^{3}$, and $\bar{v}_{270}$ is the average DM dispersion velocity in units of 270 km/s. $m_{N_i}$ denotes an element's nucleus mas in GeV. $\sigma_{i}$ is the scattering cross section off the $i$th element nucleus in pb. SI $\sigma_{i}$ is enhanced by the number of nucleons inside the nucleus and can be dominated by contributions from heavy elements, if abundant. For SD scattering, the only significant contribution in the Sun is from the hydrogen element. $f_i, F_i$ and $S$ are the mass fraction, kinematic suppression and form-factor 
suppression~\cite{Kamionkowski:1991nj} for nucleus $N_i$, respectively. $\phi_i$ describes the distributions of the $i^{th}$ element. Further details and parameter values are available in Ref.~\cite{JKG}.

We evaluate the DM annihilation rate with DarkSusy package. For the scattering cross section in this study, $\sigma<10^{-39}$cm$^2$, a minimal annihilation cross section annihilate at $\left<v\sigma\right> = 10^{-26}$ cm$^{3}$s$^{-1}$ can saturate the equilibrium condition~\cite{JKG},
\be 
\tanh^2\left(
330\left[
\frac{C}{\text{s}^{-1}}\frac{\left<v\sigma\right>}{\text{cm}^3 \text{s}^{-1}}\left(\frac{m_{\text{DM}}}{\text{10 GeV}}\right)^{0.75}
\right]^{\frac{1}{2}}
\right)\approx 1.
\ee

After annihilation, the RH neutrino $N$ leaves the Sun at a relativistic speed, and the $N$ decay produces a signal flux that is mostly along the radial direction due to the high Lorentz boost. At a distance $r$ from the center of the Sun, the decay rate over a unit volume $dV\equiv 4\pi r^2dr$ is
\be 
\frac{dN}{dtdV} = \frac{\Gamma_N}{4\pi r^2 \gamma c \tau} e^{-\frac{r}{\gamma c \tau}}
\ee
where $c$ denote the speed of light, $\gamma c \tau$ is the boosted RH neutrino decay length. The flux towards the direction of the Earth is then,
\be 
\frac{d\phi_{\gamma}}{dE}= \int d \vec{r} \frac{\Gamma_N}{4\pi r^2 \gamma c \tau}e^{-\frac{r}{\gamma c \tau}}
\cdot\frac{1}{R^2(\vec{r})}\frac{dN_{\gamma}(\theta)}{dE d\Omega}
\label{eq:integrated_photon_flux}
\ee
where $R$ is the distance between the position $\vec{r}$ and the Earth, and $\frac{dN_{\gamma}(\theta)}{dE d\Omega}$ is the boosted prompt signal (in the unit of number of particles per annihilation) in the direction of the Earth, which is at an angle $\theta$ off the radial direction. 

The boosted photon and neutrino spectra are numerically simulated with PYTHIA8 package. Due to the large Lorentz boost, most of the signal intensity is orientated along the radial direction. As a consequence, the integration over the source position $\vec{r}$ only needs to take account of a small cone towards the direction of the Sun. We use a cone-size of Fermi-LAT's 1.5$\degree$ observation window, which is large enough to capture the photon flux above the Fermi-LAT's 200 MeV energy threshold and the TeV DM scale mass in our analysis.

The photons that decay in the detectable energy range were compared to the solar gamma ray data in Ref.~\cite{fermiSun}. With a null-signal assumption, a likelihood can be calculated as,
\be
\chi^2 = \sum_i \frac{(\phi_i^{th}-\phi_i^{obs})^2}{\delta^2 \phi_i} \approx \left.\frac{(\phi^{DM})^2}{\delta^2 \phi}\right|_{\text{last bin}}
\ee
where the $\delta \phi$ is the uncertainty in the observed gamma ray flux in ~\cite{fermiSun}. As is evident in Fig.~\ref{fig:spectra}, the very hard shape in the signal photon spectrum lets the last bin completely dominate $\chi^2$. It is thus desirable to exclude the lower-energy bins in Fermi-LAT's solar gamma ray data, and only use that of the last bin in the constraint of the photon signal rate.

First a background spectrum can be obtained by fitting to Fermi-LAT's data. Focusing on the power-law shape in the relatively high energy part of the data, the best-fit background flux is 
\bea 
E^2\frac{d\phi}{dE} &=& 10^{-4.45 - 0.25 x}\ \text{MeV cm}^{-2}\text{s}^{-1}, \nn \\
\text{where } x&\equiv& \log_{10}\left(\frac{E}{\text{GeV}}\right).
\label{eq:fermi_bkg}
\eea
Due to the quickly falling power-law background spectrum, the low energy bins barely contribute to the fit of DM signal significance. To maximize the strength of constraint, only the last bin is used to fit the signal, and the 90\% confidence level corresponds to
\begin{align}
\frac{S+B-\mathrm{Fermi}}{\delta \phi} = 1.64,
\end{align}
in the highest energy bin. $B=12\times 10^{-6}\ \mathrm{MeV}\ \mathrm{cm}^{-2}\ \mathrm{s}^{-1}$ is the background from the fit in Eq.~\ref{eq:fermi_bkg}, and the measured $E^2$-flux is $ 17\times 10^{-6}\ \mathrm{MeV}\ \mathrm{cm}^{-2}\ \mathrm{s}^{-1}$. The corresponding 1$\sigma$ uncertainty is $\delta \phi = 3.4\times 10^{-6}\ \mathrm{MeV}\ \mathrm{cm}^{-2}\ \mathrm{s}^{-1}$. Note that the best-fit background is slightly lower than the measured data that makes the constraint more conservative than using the measured flux directly as the background. However, due to the very visible fluctuations in the high energy bins of Fermi-LAT's data, flux measurement variation of the order ${\cal O}(1)$ can be expected, which can be improved by enhanced statistics in future data and/or calibration with other experiments at high photon energies.

\section{neutrino signal flux}
\label{app:neutrino}

For the neutrino signal, we focus on the delayed $N$ decay outside the Sun and ignore attenuation effects inside the Sun. For interested readers, neutrino propagation effects and IceCube signal simulations are discussed in detail in Ref.~\cite{Barger:2011em}.

The neutrino source intensity calculation is similar to that of the photon signal, as is given in Eq.~\ref{eq:integrated_photon_flux}. However, due to neutrino oscillations, the relative strength of flux between different neutrino flavors changes over the distance neutrinos propagate through vacuum. For coherent oscillation, the final flavor composition after propagation over a distance $L$ is given by
\bea 
\left|\nu_i(L)\right> &=&\sum_j \left(e^{-i \hat{H} L/c}\right)_{ij}  \left|\nu_j(0)\right>\\
&\equiv& \sum_j {\cal M}_{ij} \left|\nu_j(0)\right>
\eea
where ${\cal M}$ describes the rotation in flavor and the vacuum oscillation Hamiltonian is 
\be 
{\bf H} = \frac{1}{E_\nu} {\bf V} \text{\bf diag}(0, \delta m^2_{21}, \delta m^2_{31}) {\bf V}^\dagger
\ee
and we take the latest neutrino mass-square differences $\delta m^2_{ij}$ and mixing parameters in the rotation matrix ${\bf V}$ from Ref.~\cite{nuData}. Eq.~\ref{eq:integrated_photon_flux} can be rewritten as,
\be 
\frac{d\phi_{\nu_i}}{dE}= \int d \vec{r} \frac{\Gamma_N}{4\pi r^2 \gamma c \tau}e^{-\frac{r}{\gamma c \tau}}
\cdot\frac{1}{R^2(\vec{r})} \sum_{j}|{\cal M}(R)_{ij}|^2\frac{dN_{\nu_j}(\theta)}{dE d\Omega}
\label{eq:integrated_neutrino}
\ee

At energies above few hundred GeV, vacuum oscillation length due to the solar mass splitting becomes comparable to the distance variation between the Sun and the Earth, which is due to the Earth orbit's eccentricity. As the result, oscillatory patterns in the high energy part of the neutrino spectra are still visible as shown in Fig.~\ref{fig:spectra}, after the distance average.

IceCube constraints solar DM signals by measuring up-going muon-track events in the direction of the Sun. For discussions of muon event rates and atmospheric background calculations, see Ref.~\cite{Barger:2011em} and references wherein. A full analysis would require updated knowledge of energy and angular variation (towards the Sun) in the detector fiducial volume. For this analysis, we utilize IceCube's published constraint and adopt a simplified approach, that we place the limit on the RH neutrino induced signal strength by comparing the number of integrated muon track events to those from  DM annihilation channels into $\tau^\pm$ and $W^\pm$ final states, or
\be 
\sigma = \sigma_{\tau^\pm} \frac{N_\mu}{N_\mu^{\tau^\pm}}.
\label{eq:scaling}
\ee
where the current bound on $\sigma_{\tau^\pm}$ is given in Ref.~\cite{ICBounds}.

The muon rate from $N$ decay is calculated from the muon neutrino flux with the GENIE~\cite{genie} package. 
For various DM masses $M_{DM}$, muon event number is integrated starting from an optimized threshold energy, above which the signal strength is large enough to give statistic significance against the atmospheric background. For a few DM masses $M_{DM}$= \{200, 500, 1000, 2000, 5000\} GeV, 
the respective threshold energy for contained muons are \{75, 100, 150, 175, 200\} GeV. These energy thresholds depend on the shape of the signal spectrum and can vary for different annihilation final states. The muon event rate for DM + DM $\rightarrow NN$ is then compared to that from a DM + DM $\rightarrow \tau^+\tau^-$. We then fold the ratio of the muon events for these final states into Eq.~\ref{eq:scaling} to derive the corresponding constraint on the DM-nucleon elastic scattering cross section in the case of DM annihilation into $N$. Admittedly, this is approximation to a fully experimentally-simulated $\tau^+\tau^-$ channel's muon event rate in the direction of the Sun, yet we argue that the muon event rate ratio(s) between different signal channels are not sensitive to the details of experimental setup for high (TeV) neutrino energy much above IceCube's measurement threshold that is less than 70 GeV. We used WimpSim~\cite{wimpsim} for the neutrino fluxes in $\tau^\pm$ and $W^\pm$ channels.


\begin{thebibliography}{99}


 \bibitem{bib:sun} 
  Y.~B.~Zeldovich, A.~A.~Klypin, M.~Y.~Khlopov and V.~M.~Chechetkin,
  Sov.\ J.\ Nucl.\ Phys.\  {\bf 31} (1980) 664
  [Yad.\ Fiz.\  {\bf 31} (1980) 1286].
  J.~Silk, K.~A.~Olive and M.~Srednicki,
  Phys.\ Rev.\ Lett.\  {\bf 55}, 257 (1985).
  L.~M.~Krauss, K.~Freese, W.~Press and D.~Spergel,
  Astrophys.\ J.\  {\bf 299}, 1001 (1985).
  T.~K.~Gaisser, G.~Steigman and S.~Tilav,
  Phys.\ Rev.\  D {\bf 34}, 2206 (1986).
  L.~M.~Krauss, M.~Srednicki and F.~Wilczek,
  Phys.\ Rev.\  D {\bf 33}, 2079 (1986).

\bibitem{Adrian-Martinez:2016gti} 
  S.~Adrian-Martinez {\it et al.} [ANTARES Collaboration],
  Phys.\ Lett.\ B {\bf 759}, 69 (2016)
  doi:10.1016/j.physletb.2016.05.019

\bibitem{Low}
C. Rott, J. Siegal-Gaskins and J. F. Beacom, Phys. Rev. D {\bf 88}, 055005 (2013) [arXiv:1208.0827 [astro-ph.HE]]; N. Bernal, J. Martin-Albo and S. Palomares-Ruiz, JCAP 1308, 011 (2013) [arXiv:1208.0834 [hep-ph]].


\bibitem{JKG}
G. Jungman, M. Kamionkowski and K. Griest, Phys. Rept. {\bf 267}, 195 (1996) [hep-ph/9506380]. 


\bibitem{ICBounds}
M. G. Aartsen {\it et al.} [IceCube Collaboration], JCAP {\bf 1604}, 022 (2016) [arXiv:1601.00653 [hep-ph]].


\bibitem{LUX}
D. S. Akerib {\it et al.} [LUX Collaboration], arXiv:1608.07648 [astro-ph.CO]. 


\bibitem{PandaX}
A.  Tan {\it et  al.} [PandaX Collaboration], arXiv:1607.07400 [hep-ex].


\bibitem{Schuster:2009au} 
  P.~Schuster, N.~Toro and I.~Yavin,
  Phys.\ Rev.\ D {\bf 81}, 016002 (2010)

\bibitem{Batell:2009zp} 
  B.~Batell, M.~Pospelov, A.~Ritz and Y.~Shang,
  Phys.\ Rev.\ D {\bf 81}, 075004 (2010)

\bibitem{Menon:2009qj} 
  A.~Menon, R.~Morris, A.~Pierce and N.~Weiner,
  Phys.\ Rev.\ D {\bf 82}, 015011 (2010)

\bibitem{Berger:2014sqa} 
  J.~Berger, Y.~Cui and Y.~Zhao,
  JCAP {\bf 1502}, no. 02, 005 (2015)


\bibitem{Meade:2009mu} 
  P.~Meade, S.~Nussinov, M.~Papucci and T.~Volansky,
  JHEP {\bf 1006}, 029 (2010)


\bibitem{MM}
R. N. Mohapatra and R. E. Marshak, Phys. Rev. Lett. {\bf 44}, 1316 (1980) [Erratum-ibid. 44, 1643 (1980)].


\bibitem{ABDR}
R. Allahverdi, S. Bornhauser, B. Dutta and K. Richardson-McDaniel, Phys. Rev. D {\bf 80}, 055026 (2009) 


\bibitem{ACD}
R. Allahverdi, S. Campbell and B. Dutta, Phys. Rev. D {\bf 85}, 035004 (2012) 


\bibitem{ACDG} 
R. Allahverdi, S. S. Campbell, B. Dutta and Y. Gao, Phys. Rev. D {\bf 90}, 073002 (2014) 

\bibitem{bib:typeI}
P.~Minkowski, Phys. Lett. B {\bf 67}, 421 (1977);
%
T.~Yanagida, in \emph{Proceedings of the Workshop on the Unified
  Theory and the Baryon Number in the Universe} (O.~Sawada and
  A.~Sugamoto, eds.), KEK, Tsukuba, Japan, 1979, p.~95;
%
M.~Gell-Mann, P.~Ramond, and R.~Slansky, \emph{Supergravity} (P.~van
  Nieuwenhuizen et al. eds.), North Holland, Amsterdam, 1979, p.~315;
%
S.~L. Glashow, \emph{The future of elementary particle physics}, in
  \emph{Proceedings of the 1979 Carg{\`e}se Summer Institute
 on Quarks and Leptons} (M.~Levy et al. eds.),
 Plenum Press, New York, 1980, p.~687;
%
R.~N. Mohapatra and G.~Senjanovic,
 Phys. Rev. Lett. {\bf 44}, 912 (1980).


\bibitem{split}
A. Kusenko, F. Takahashi and T. T. Yanagida, Phys. Lett. B {\bf 693}, 144 (2010) 


\bibitem{LHC}
A. Das and N. Okada, Phys. Rev. D {\bf 88}, 113001 (2013) 113001 [arXiv:1207.3734 [hep-ph]]; J. C. Helo, M. Hirsch and S. Kovalenko, Phys. Rev. D {\bf 89}, 073005 (2014) [arXiv:1312.2900 [hep-ph]]; P. S. B. Dev, A. Pilaftsis and U.-k. Yang, Phys. Rev. Lett. {\bf 112}, 081801 (2014) [arXiv:1308.2209 [hep-ph]];
A. Das, P. S. Bhupal Dev and N. Okada, Phys. Lett. B {\bf 735}, 364 (2014) [arXiv:1405.0177 [hep-ph]]; J. N. Ng, A. de la Puente and B. W.-P. Pan, JHEP {\bf 1512}, 172 (2015) [arXiv:1505.01934 [hep-ph]]; E. Izaguirre and B. Shuve, Phys. Rev. D {\bf 91}, 093010 (2015) [arXiv:1504.02470 [hep-ph]]; P. Q. Hung, T. Le, V. Q. Tran and T.-C. Yuan, JHEP {\bf 1512}, 169 (2015) [arXiv:1508.07016 [hep-ph]]; T. Peng, M. J. Ramsey-Musolf and P. Winslow, Phys. Rev. D {\bf 93}, 093002 (2016) [arXiv:1508.04444 [hep-ph]]; P. S. B. Dev, D. Kim and R. N. Mohapatra, JHEP {\bf 1601}, 118 (2016) [arXiv:1510.04328 [hep-ph]];
J. Gluza and T. Jelinski, Phys. Lett. B {\bf 748}, 125 (2015) [arXiv:1504.05568 [hep-ph]]; A. M. Gago, P. Hernández, J. Jones-Pérez, M. Losada and A. Moreno Briceño, Eur. Phys. J. C {\bf 75}, 470 (2015) [arXiv:1505.05880 [hep-ph]]; A. Das and N. Okada, Phys. Rev. D {\bf 93}, 03303 (2016) [arXiv:1510.04790 [hep-ph]]; Z. Kang, P. Ko and J. Li, Phys. Rev. D {\bf 93}, 075037 (2016) [arXiv:1512.08373 [hep-ph]]; C. O. Dib and C. S. Kim, Phys. Rev. D {\bf 92}, 093009 (2015) [arXiv:1509.05981 [hep-ph]]; C. Degrande, O. Mattelaer, R. Ruiz and J. Turner, arXiv:1602.06957 [hep-ph]; L. Duarte, J. Peressutti and O. A. Sampayo, arXiv:1610.03894. 


\bibitem{LC}
A. Das and N. Okada, Phys. Rev. D {\bf 88}, 113001 (2013) [arXiv:1207.3734 [hep-ph]]; T. Asaka and T. Tsuyuki, Phys. Rev. D {\bf 92}, 094012 (2015) [arXiv:1508.04937 [hep-ph]]; S. Banerjee, P. S. B. Dev, A. Ibarra, T. Mandal and M. Mitra, Phys. Rev. D {\bf 92}, 075002 (2015) [arXiv:1503.05491 [hep-ph]]; A. Blondel, E. Graverini, N. Serra and M. Shaposhnikov, in Proceedings, 37th International Conference on High Energy Physics (ICHEP 2014) [arXiv:1411.5230 [hep-ph]]; S. Antusch, E. Cazzato and O. Fischer, JHEP {\bf 1604}, 189 (2016) [arXiv:1512.06035 [hep-ph]]; A. Abada, D. Becirevic, M. Lucente and O. Sumensari, Phys. Rev. D {\bf 91}, 113013 (2015) [arXiv:1503.04159 [hep-ph]]; S. Antusch and O. Fischer, JHEP {\bf 1505}, 053 (2015) [arXiv:1502.05915 [hep-ph]]; S. Antusch and O. Fischer, Int. J. Mod. Phys. A {\bf 30}, 1544004 (2015).


\bibitem{Meson}
D. Milanes, N. Quintero and C. E. Vera, Phys. Rev. D {\bf 93}, 094026 (2016);
G. Cvetic and C. S. Kim, arXiv:1606.04140 [hep-ph]; T. Asaka and H. Ishida, arXiv:1609.06113 [hep-ph].


\bibitem{FT}
D. S. Gorbunov and M. E. Shaposhnikov, JHEP {\bf 0710}, 015 (2007); 
S. N. Gninenko, D. S. Gorbunov and M. E. Shaposhnikov, Adv. High Energy Phys. {\bf 2012}, 718259 (2012);
T. Asaka, S. Eijima and A. Watanabe, JHEP {\bf 1303}, 125 (2013) 


\bibitem{SHIP}
B. Batell, M. Pospelov and B. Shuve, JHEP {\bf 1608}, 052 (2016) 


\bibitem{0nu2beta}
M. Drewes and S. Eijima, arXiv:1606.06221 [hep-ph]; T. Asaka, S. Eijima and H. Ishida, arXiv: 1606.06686 [hep-ph].

\bibitem{bib:pdg2016}
C. Patrignani {\it et al.} (Particle Data Group), Chin. Phys. C {\bf 40}, 100001 (2016).

\bibitem{bib:cosmo}
 P.~A.~R.~Ade {\it et al.} [Planck Collaboration],
  Astron.\ Astrophys.\  {\bf 594}, A13 (2016)
  doi:10.1051/0004-6361/201525830
For a recent anaylsis, see M.~M.~Zhao, Y.~H.~Li and X.~Zhang,
  arXiv:1608.01219 [astro-ph.CO].


\bibitem{Decay}
M. Dittmar, A. Santamaria, M. C. Gonzalez-Garcia and J. W. F. Valle, Nucl. Phys. B {\bf 332}, 1 (1990); M. C. Gonzalez-Garcia, A. Santamaria and J. W. F. Valle, Nucl. Phys. B {\bf 342}, 108 (1990).


\bibitem{Excess}
Y-L Tang and S-h Zhu, JHEP {\bf 1603}, 043 (2016) 


\bibitem{Rius}
M. Escudero, N Rius, and V Sanz, arXiv:1606.01258 [hep-ph]; M. Escudero, N. Rius and V. Sanz, arXiv:1607.02373 [hep-ph]. 


\bibitem{feynrules}
  A. Alloul, N. D. Christensen, C. Degrande, C. Duhr and B. Fuks,
  Computer Physics Communications {\bf 185}, 2250 (2014) 


\bibitem{bib:ib}
  V.~Barger, Y.~Gao, W.~Y.~Keung and D.~Marfatia,
  Phys.\ Rev.\ D {\bf 80}, 063537 (2009)
[arXiv:0906.3009 [hep-ph]].
  
  G.~Bambhaniya, J.~Kumar, D.~Marfatia, A.~C.~Nayak and G.~Tomar,
  arXiv:1609.05369 [hep-ph].


\bibitem{pythia8}
  T. Sj{\"o}strand, S. Mrenna and P. Skands,
  Computer Physics Communications {\bf 178}, 852 (2008) [arXiv:0710.3820 [hep-ph]].

\bibitem{darksusy}
  P. Gondolo, J. Edsj{\"o}, P. Ullio, L. Bergstr{\"o}m, M. Schelke and E. A. Baltz,
  JCAP {\bf 0407}, 8 (2004) [arXiv:astro-ph/0406204].

\bibitem{atmosphericBackground}
  M. Honda, M. S. Athar, T. Kajita, K. Kasahara and S. Midorikawa,
  Phys. Rev. D {\bf 92}, 023004 (2015) [arXiv:1502.03916 [astro-ph]].


\bibitem{fermiSun}
  K. C. Y. Ng, J. F. Beacom, A. H. G. Peter and C. Rott,
  Phys. Rev. D {\bf 94}, 023004 (2016) [arXiv:1508.06276 [astro-ph]].


\bibitem{nuData}
  D. V. Forero, M. T{\'o}rtola and J. W. F. Valle,
  Phys. Rev. D {\bf 90}, 093006 (2014) [arXiv:1405.7540 [hep-ph]].


\bibitem{pico}
  C. Amole, M. Ardid, D. M. Asner, D. Baxter, E. Behnke, P. Bhattacharjee et al.,
  Phys. Rev. D {\bf 93}, 052014 (2016) [arXiv:1510.07754 [hep-ex]].


\bibitem{HAWC}
M. L. Proper {\it et al.} [HAWC Collaboration], PoS ICRC2015, 1213 (2016) 1213 [arXiv:1508.04470 [astro-ph.HE]].

\bibitem{Feng:2014vza} 
  C.~Feng {\it et al.} [DAMPE Collaboration],
  arXiv: arXiv:1406.3886 [astro-ph.IM].

\bibitem{Kamionkowski:1991nj}
  M.~Kamionkowski,
  Phys.\ Rev.\  D {\bf 44}, 3021 (1991).

\bibitem{Barger:2011em} 
  V.~Barger, Y.~Gao and D.~Marfatia,
  Phys.\ Rev.\ D {\bf 83}, 055012 (2011)
[arXiv:1101.4410 [hep-ph]].

\bibitem{genie}
  C. Andreopoulos, A. Bell, D. Bhattacharya, F. Cavanna, J. Dobson, S. Dytman et al.,
  Nuclear Instruments and Methods in Physics Research A {\bf 614}, 87 (2010) 
  [arXiv:0905.2517 [hep-ph]].

\bibitem{wimpsim}
  M. Blennow, J. Edsj{\"o} and T. Ohlsson,
  JCAP {\bf 0801}, 21 (2008) [arXiv:0709.3898 [hep-ph]].

\end{thebibliography}
\end{document}